\documentclass[trackchanges,twocolumn]{aastex7}

\usepackage{graphicx}
\usepackage{amsmath}
\usepackage{natbib}
\usepackage[T1]{fontenc}
\usepackage{float}
\usepackage{txfonts}
\begin{document}

    \title{Accretion onto Reissner-Nordstr\"{o}m naked singularities}

    \author[0000-0002-1982-0944]{Tomasz Krajewski}
    \affiliation{Nicolaus Copernicus Astronomical Center of the Polish Academy of Sciences, Bartycka 18, 00-716 Warsaw, Poland}
    \affiliation{Institute of Fundamental Technological Research of the Polish Academy of Sciences, Pawi\'nskiego 5B, 02-106 Warsaw, Poland}
    \email{tkrajewski@camk.edu.pl}
    
    \author[0000-0001-9043-8062]{W\l{}odek Klu\'{z}niak}
    \affiliation{Nicolaus Copernicus Astronomical Center of the Polish Academy of Sciences, Bartycka 18, 00-716 Warsaw, Poland}
    \email{wlodek@camk.edu.pl}
    

\makeatletter
\renewcommand{\frontmatter@title@above}{}
\makeatother

    \begin{abstract}
   Nearly every galactic core contains a supermassive compact object, hypothesized to be a Kerr black hole. It was only with the advent of Event Horizon Telescope observations that the predictions of this hypothesis could be observationally tested for our own Galaxy, and the nearby elliptical M87, on spatial scales comparable to the gravitational radius. At the same time it became possible to test whether alternatives such as naked singularities in general relativity, or similar objects in alternative theories of gravity, are excluded by the data. These and other observational developments renewed interest in non-Kerr spacetime metrics, also in the context of active galactic nuclei at cosmological distances.
   Recently, we have shown that accreting naked singularities in the Reissner-Nordstr\"{o}m metric of general relativity tend to produce strong outflows. The geometry and origin of these winds is studied here, and their parameter dependence is investigated. To this end we performed numerical GR hydrodynamical simulations of accretion of electrically neutral matter in the Reissner-Nordstr\"{o}m metric and discussed the results in the context of analytic predictions of fluid motion in this spacetime.
    \end{abstract}



\section{Introduction}
We investigate the process of accretion onto a naked singularity described by the Reissner-Nordstr\"{o}m (RN) metric in the regime when the charge $Q$ is slightly larger in magnitude than the mass $M$ of the compact object. For $|Q| \le M$ the RN metric describes the gravitational field of an (electrically) charged black hole with an~event horizon. The interior of the black hole, i.e. the part of the spacetime inside the horizon, is hidden from distant observers. A black hole usually has a~singularity located close to its center, well below the horizon, but regular solutions have also been considered in the literature \citet{Bardeen:1968,Ayon-Beato:1998hmi,Hayward:2005gi,Frolov:2016pav}. 

For $|Q| > M$ the RN metric does not have an~event horizon, thus the singularity is directly visible to distant observers. Such objects are called naked singularities. The problem of causality in naked singularity (NkS) spacetimes led to the cosmic censorship conjecture which would exclude their existence \citep{Penrose69}. However, it is not even clear how to formulate the cosmic censorship hypothesis \citep{Joshi93}, seeing how a naked singularity has been recognized in an exact solution of Einstein's equations representing a contracting radiating star \citep{SteinKingLasota:1975}, and several calculations followed a gravitational collapse of dust \citep{eardleySmarr1979,christodoulou1984}, and perfect fluids \citep{oriPiran1990,giamboGiannoniMagliPiccione2004} to a naked singularity.

As we will argue later there is a~characteristic length scale called zero-gravity radius of spherically symmetric naked singularity spacetimes that is crucial for the process of accretion, in the same sense as the horizon radius is important for accretion onto black holes. We can consider the situation when this scale is macroscopic and the problem of mathematical singularity is resolved at microscopic scales as it is in the case of regular black holes, thus avoiding problems with causality. From the perspective of effective field theory, General Relativity (GR) is a low energy effective theory which describes the gravitational interactions at large (macroscopic) length scales, and the singularities found in solutions of Einstein's equations should be cured in the UV completion, especially by effects of quantum gravity. Even though the problem of interactions of subatomic naked singularities driven by quantum mechanics seems to be an extremely exciting topic, it is far beyond the scope of this paper and will be postponed to future studies.

The zero-gravity radius is the radius of the so called zero-gravity sphere, i.e. a sphere on which a test particle can stay at rest. The spacetime inside the zero-gravity sphere, i.e. at distances from the central (point-like) singularity smaller than the zero-gravity radius, has an intriguing property: test particles are expelled from this region and gravity is effectively repulsive. This feature of the RN metric allows for the formation of a levitating atmosphere \citep{Vieira:2023cvn}. Furthermore, as was shown in \citet{Mishra:2024bpl}, the repulsive nature of the gravitational field in the vicinity of naked singularities influences the shape of (toroidal) figures of equilibrium in the background of the RN metric.

We expect accretion onto spherically symmetric naked singularities in modified theories of gravity to proceed in a manner qualitatively similar to the one onto a naked singularity described by the RN metric. It is known that various modified theories of gravity support naked singularity solutions with zero-gravity spheres similar to the RN one \citep{Vieira:2023cvn, ruchiphd}. Furthermore, even the RN metric is a~solution of equations of certain modified theories of gravity for which the parameter $Q$ does not correspond to the electric charge \citep{ruchiphd}. These include BBMB theory \citep{Bocharova:1970, Bekenstein:1974sf}, Horndeski gravity \citep{Babichev:2017guv}, Randall-Sundrum II braneworld "black hole solution" \citep{Aliev:2005bi}, and Moffat's modified gravity \citep{Moffat:2014aja}. Even in non-spherically symmetric spacetimes, such as that of mildly rotating ($1<a_*\approx1$) Kerr NkS \citep{DihingiaAkhil25} there is some resemblance to the RN NkS results

In this work we are interested in objects of astrophysical sizes, i.e. with mass, $M$, on the order of masses of planets, stars or even supermassive compact objects in centers of galaxies. In this regime the zero-gravity sphere would be of macroscopic size, for which GR was already extensively tested \citet{Adelberger:2001,Hoyle:2004cw,Tan:2020vpf,Lee:2020zjt,Westphal:2020okx,Blakemore:2021zna,Fuchs:2023ajk}. 
allowing us to use GR hydrodynamics in order to describe motion of fluid in these spacetimes. Moreover, we will neglect self-gravity of the accreted matter, assuming that the mass involved in the process is much smaller than the mass of the central object.

In our studies we used the \texttt{Koral+} code, which is an~extension of the well-known \texttt{Koral} code described by \cite{Sadowski:2012, Sadowski:2013gua}. In the version of the code that we used, the metric dependent part was refactorized and a new one was added that simplifies implementation of new systems of coordinates by exploiting symbolic computations software. With this numerical tool we investigated the accretion of electrically neutral matter in the RN metric background.

Accretion onto naked singularities described by the RN metric with $ 1 < Q/M < \sqrt{5}/2$ is covered in the current paper. In this regime it is possible to construct equilibrium tori with a cusp at the self-intersection of a critical equipotential surface, the cusp allows dynamical accretion to proceed through the inner edge of the torus.\footnote{The inflow of matter through the cusp is possible without loss of angular momentum, so we were able to perform purely hydrodynamical simulations of accretion with no viscosity.}

The paper is organized as follows. In Section~\ref{sec:stability_analysis} we briefly summarize the important facts about structure of the spacetime described by the RN metric. Section~\ref{sec:equilibrium_tori} contains a brief discussion of equilibrium tori in the RN metric that we used as initial conditions for our simulations. Section~\ref{sec:sims} presents main characteristics of our numerical simulations, while the description of the numerical techniques used in our studies is postponed to Appendix~\ref{sec:numerical_code}. In Section \ref{sec:numerical_results} results of our numerical simulations of accretion onto naked singularities are presented. We summarize and conclude in Section~\ref{sec:summary}. In Appendix~\ref{sec:MBO_roots} we present exact analytic expressions for location of marginally bound orbits in the RN metric. Appendix~\ref{sec:supplemental} contains additional results from our numerical simulations which are not needed to understand the results discussed in the main text of the paper, but may be useful for other researches working in the field.

\section{Reissner-Nordström naked singularity\label{sec:stability_analysis}}
\begin{figure}[!]
\centering 
\includegraphics[width=\columnwidth]{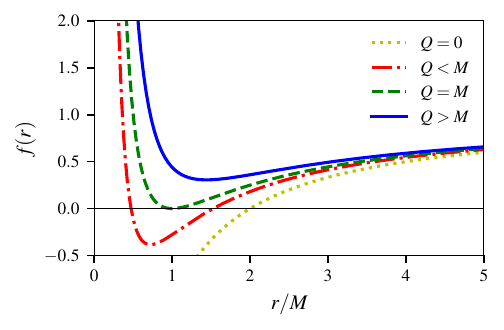}
\caption{The time-time component of the Reissner-Nordström metric as a function of the radius
for various values of the charge to mass ratio $Q/M$. Blue solid line: $-g_{tt} = f(r)$ for naked singularity with $Q / M = 1.2$. Green dashed line: extremal black hole $Q = M$, for which only one horizon exists. Red dotted-dashed line: black hole with $Q / M= 0.85$. For comparison, $-g_{tt}$ of Schwarzschild black-hole metric ($Q=0$) is plotted as dotted yellow line.
}
\label{fig:gtt_plot}
\end{figure}
The static metric describing the gravitational field set by a charged,\footnote{In fact
$Q$ does not need to be the electric charge and may be charge of any (hypothetical) Abelian gauge symmetry.} spherically symmetric, body of mass M without spin is given by the metric tensor field
\begin{equation}
g=-f(r)c^2dt^2+\frac{1}{f(r)}dr^2+r^2(d\theta^2+\sin^2\theta^2d\phi^2).
\label{eq:Reissner_Nordstrom_metric}
\end{equation}
We use spherical coordinates $(t,r,\theta,\phi)$, where $t$ is the time coordinate measured by a stationary clock at infinity. The metric \eqref{eq:Reissner_Nordstrom_metric} is a solution of Einstein-Maxwell set of equations for the metric function $f(r)=1-2r_{g}/r+r_{Q}^2/r^2$, where in the $G=c=1$ unit convention the gravitational radius of the body of mass $M$ is defined as $r_g=M$, the gravitational time as $t_g=M$, and the characteristic length scale of charge as $r_Q=|Q|$.

In the case with no charge, $Q=0$, we obtain the Schwarzschild metric with $f(r)=1-2r_g/r$, which describes a black hole with an event horizon at radius $r_h = 2r_g$.

There are two horizons when $0 < Q < M$, the outer and the inner one, located at $r_h^+$ and $r_h^-$ respectively, corresponding to two zeros of the metric function $f(r)$:
\begin{equation}
r_h^\pm = r_g \pm \sqrt{r_g^2 - r_Q^2} = M \pm \sqrt{M^2 - Q^2}.
\end{equation}
For $Q=M$ both horizons coincide. For $Q^2 > M^2$ the metric function $f$ is always strictly positive, thus the solution does not have an event horizon at all. This is the naked singularity solution.

\begin{figure}[!]
\centering 
\includegraphics[width=\columnwidth]{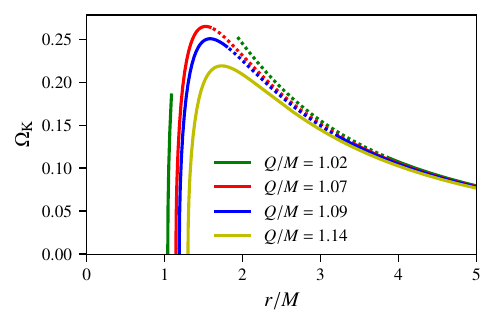}
\caption{Radial dependence of Keplerian frequency, $\Omega_\mathrm{K}$, in circular orbits for Reissner-Nordström NkS with charge $Q$ and mass $M$,
 for (top to bottom) $Q/M = 1.02$, $1.07$, $1.09$, $1.14$. Solid lines represent stable orbits, dotted lines represent unstable orbits. 
The green curve is disrupted by the forbidden region between the photon orbits (c.f. Fig.~\ref{fig:stability_plot}).
}
\label{fig:omega_plot}
\end{figure}

Even though the Reissner-Nordström metric does not have a horizon in the regime $|Q| > M$, there is a characteristic length scale $r_0$ which is crucial for the accretion onto naked singularities. It turns out that our weak-field limit intuitions of gravity that is always attractive fails in the case of naked singularities. Unlike for Kerr (or Schwarzschild) and RN black hole metrics, which only allow test particles to stay at rest asymptotically at infinite distance, the RN naked singularity metric features a so called zero-gravity sphere at which the acceleration of a test particle vanishes, as first described by \cite{Pugliese:2010ps}. Its radius, the zero-gravity radius, is given by a root of the derivative of metric function $f'(r_0) = 0$, specifically, 
\begin{equation}
    r_0 = Q^2/M
\end{equation}
for the RN metric. Interestingly, the charge lengthscale is the geometric mean of the gravitational radius and the zero-gravity radius,
$r_Q=\sqrt{r_0r_g}$. The zero-gravity radius plays a~crucial role for naked singularities. For instance, inside the zero gravity sphere, i.e. at $r<r_0$, gravity is effectively repulsive. It is this feature of the RN metric that allows the formation of the non-rotating levitating atmospheres, and determines shapes of the toroidal figures of equilibrium.

Another interesting feature of the RN metric related to the zero-gravity sphere is the unusual radial dependence of Keplerian frequency near the NkS.
By evaluating the frequency of motion in circular orbits through $\Omega_\mathrm{K}(r) =\sqrt{f'(r)/(2r)}$, we obtain
\begin{equation}
    \Omega_\mathrm{K}(r) = \sqrt{\frac{M}{r^3}\left(1 - \frac{Q^2}{Mr}\right)}=\sqrt{\frac{M}{r^3}}\sqrt{1-\frac{r_0}{r}}.
    \label{eq:keplerian}
\end{equation}

Unlike in the case of the Kerr metric,\footnote{In the Kerr metric \mbox{$\Omega_\mathrm{K} = \pm M^{1/2} / \left(r^{{3}/{2}} \pm a M^{{1}/{2}}\right)$}, where $a$ is the spin of the source of gravitational field and the $+(-)$ sign is for prograde (retrograde) orbits.} this Keplerian frequency has a maximum. Indeed, while equation~(\ref{eq:keplerian}) has the usual Schwarzschild (and Newtonian) limit for $r>>r_0$, and tends to zero at infinity, the attenuating factor $\sqrt{1-r_0/r}$ reduces $\Omega_\mathrm{K}$ smoothly to zero at $r=r_0$, so the orbital frequency must have a maximum between $r_0$ and infinity. The maximum of $\Omega_\mathrm{K}$ turns out to be at
\begin{equation}
    r_{\Omega \text{max}}=(4/3)\,r_0,
    \label{eq:rOmegamax}
\end{equation} 
fairly close to the zero-gravity sphere. The radial dependence of $\Omega_\mathrm{K}$ for the RN metric corresponding to naked singularities is presented in Fig.~\ref{fig:omega_plot}.

\begin{figure}[!]
\centering 
\includegraphics[width=\columnwidth]{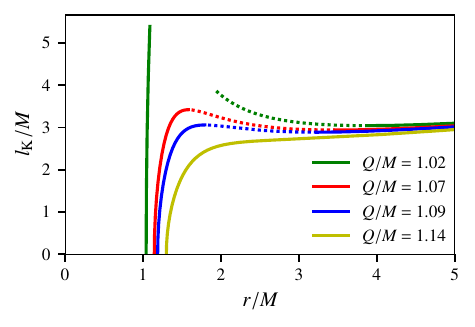}
\caption{Keplerian angular momentum, $l_\mathrm{K}$, of circular orbits for Reissner-Nordström NkS, with (top to bottom) $Q/M = 1.02$, $1.07$, $1.09$, $1.14$. Solid lines represent stable orbits and dotted ones unstable orbits. The green curve is disrupted by the forbidden region between photon orbits.
Circular orbits are stable when they satisfy the Rayleigh criterion, $dl_\mathrm{K}/dr>0$.}
\label{fig:ell_plot}
\end{figure}

Owing to the spherical symmetry of the RN metric, the specific angular momentum $l$ of a test particle is conserved during its motion. The dependence of the Keplerian specific angular momentum $l=-u_\phi/u_t$ on the radius of the circular orbit is given as:
\begin{equation}
    l_\mathrm{K} = r^2 f(r)^{-1} \Omega_\mathrm{K}(r) = \frac{\sqrt{M r - Q^2}}{1 - \frac{2M}{r} + \frac{Q^2}{r^2}},
\end{equation}
and presented in Fig.~\ref{fig:ell_plot}.

\begin{figure}[!]
\centering 
\includegraphics[width=\columnwidth]{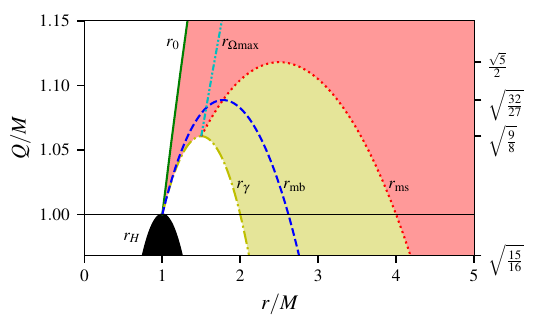}
\caption{Stability of circular orbits in Reissner-Nordström metric with charge $Q$ and mass $M$. Black patch corresponds to region between horizons. Solid green line: the zero-gravity radius. Dashed dotted yellow line: the radii of photon orbits. Dotted red line: radii of marginally stable orbits. Blue dashed line: radii of marginally bound orbits. Cyan line: location of the maximum of Keplerian frequency of circular orbits. Stable circular timelike geodesics exist in the shaded red region. In the shaded yellow region (sandwiched between the red and white ones) only unstable circular timelike geodesics exist.}
\label{fig:stability_plot}
\end{figure}

Other important quantities describing the spacetime of the RN metric are locations of circular photon orbits, and the marginally stable and marginally bound orbits of (time-like) test particles. The former two were investigated by \cite{Pugliese:2010ps}, who showed that the radii of circular photon orbits $r_\gamma$ are the roots of 
\begin{equation}
    r_\gamma^2 - 3 M r_\gamma + 2 Q^2 = 0,
\end{equation}
with two real solutions 
\mbox{$r_{\gamma \pm} = \left(3 M \pm \sqrt{9 M^2 - 8 Q^2} \right)/2$} for \mbox{$ 1 < |Q| /M < \sqrt{9/8}$}, which coincide in the limit of \mbox{$|Q| /M = \sqrt{9/8}$}, and no real roots for \mbox{$|Q| /M > \sqrt{9/8}$}.
The outer circular photon orbit at $r_{\gamma+}$ is the unstable photon orbit familiar from studies of black holes. The inner one at $r_{\gamma-}$ is stable against radial perturbations and is only present around naked singularities. 

The radius $r_{\text{ms}}$ of a marginally stable orbit is given by the root of 
\begin{equation}
    M r_{\text{ms}}^3 - 6 M^2 r_{\text{ms}}^2 + 9 M Q^2 r_{\text{ms}} - 4 Q^4 = 0.\label{eq:LSCO_equation}
\end{equation}
According to the analysis in \cite{Pugliese:2010ps} Eq. \eqref{eq:LSCO_equation} has two real roots larger than the zero-gravity radius $r_0$ for $ 1 < |Q| / M < \sqrt{5}/2$, one such real root for $|Q| / M = \sqrt{5}/2$, and no such real roots for $ |Q| / M > \sqrt{5}/2$. 

Thus, for $|Q| / M > \sqrt{5}/2$ all circular orbits outside zero-gravity radius $r_0$ are stable. In this regime the Keplerian angular momentum is a monotonic function of the radius, $dl_\mathrm{K}/dr>0$. A particle initially in a distant circular orbit will drift towards the NkS through a progression of tighter orbits, if only there is a mechanism adiabatically removing its angular momentum. It is thought that in an ionized geometrically thin accretion disk such a mechanism is operative, at least as long as $d\Omega_\mathrm{K}/dr<0$ \citep[the effective turbulent viscosity caused by the magnetorotational instability, MRI][]{Balbus:2003xh}. The angular velocity of the fluid in thin disks closely matches that of a test particle \citep{ShakuraSunyaev:1973}. For the RN NkS the thin accretion disk is then expected to extend down to about $r=4r_0/3$ (Eq.~\ref{eq:rOmegamax}). If a way were found to further remove the angular momentum of the fluid, it could settle on the zero-gravity sphere, with $\Omega_\mathrm{K}$ smoothly going to zero as $r\rightarrow r_0$. Simulations suggest that this indeed happens, at the expense of the disk becoming geometrically thick for $r\leq r_{\Omega \text{max}}$ \citep{Cemeljic:2025}.

For $|Q| / M \le \sqrt{5}/2$ circular test-particle orbits may be stable, unstable, or may not exist at all, depending on the value of $|Q|/M$ and the radius of the circular orbit.
It turns out that for $1 < |Q| /M < \sqrt{9/8}$ one of the roots of \eqref{eq:LSCO_equation} lies between the circular photon orbits, so in the radial interval $]r_{\gamma-},r_{\gamma+}[$ corresponding to a region which does not contain timelike geodesics. As a result two marginally stable orbits exist only for $\sqrt{9/8}< |Q|/M<\sqrt{5}/2$.

For $ 1 < |Q| /M \le \sqrt{9/8}$ there is only one marginally stable orbit, and there are two regions of stable time-like circular orbits, with orbital radii in the intervals $]r_0, r_{\gamma-}[$ and $[r_{\text{OMSCO}}, \infty[$, where $r_{\text{OMSCO}}$ is the largest root of \eqref{eq:LSCO_equation}, i.e. it is the radius of the outer\footnote{The words ``outer'' and ``inner'' only signify the relative radial position of the orbits. In fact the OMSCO is the inner boundary of the ordinary region of stability extending to infinity, and the EMSCO is the outer boundary of the additional inner region of stability near the zero-gravity sphere.}
marginally stable circular orbit, that is one of the largest extent. It could be called the ordinary marginally stable circular orbit (OMSCO), as it has properties quite similar to the black hole innermost circular orbit (ISCO). The inner one, which only exists for $\sqrt{9/8}<|Q|/M<\sqrt{5}/2$ could be called the EMSCO (extraordinary marginally stable circular orbit), it is unusual in the sense that a small radial perturbation of the orbit would cause the test particle to move \emph{away} from the central object.

\begin{figure}[!]
\centering 
\includegraphics[width=\columnwidth]{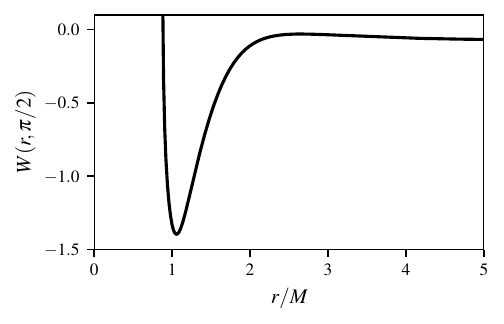}
\caption{Effective potential $W(r,\theta)$ in the equatorial plane ($\theta = \pi/2$) as a~function of the radius $r$ for Reissner-Nordström NkS metric with $Q/M = 1.02$ for a selected value of $l_0=3.22$.
}
\label{fig:W_example}
\end{figure}

The radius $r_\text{mb}$ of marginally bound orbit satisfies the following equation
\begin{equation}
    M r_\text{mb}^3 - 4 M^2 r_\text{mb}^2 + 4 M Q^2 r_\text{mb} - Q^4 = 0;
\label{eq:MBO_equation}
\end{equation}
\citep{Beheshti:2015bak}. Eq. \eqref{eq:MBO_equation} has two real roots grater than $r_0$ for $ 1 < |Q| / M < \sqrt{32 /27}$, one such real root for $|Q| / M = \sqrt{32 /27}$, and no such real roots for $|Q| / M > \sqrt{32 /27}$. Our stability analysis of circular orbits in the spacetime of the RN metric is summarized in Fig.~\ref{fig:stability_plot}.

\section{Hydrostatic tori around Reissner-Nordström naked singularity\label{sec:equilibrium_tori}}

In the following Section we will briefly summarize the findings of \cite{Mishra:2024bpl} regarding the figures of equilibrium in RN naked singularity metrics, taking a closer look at the details important for understanding the dynamics of perfect fluid in our numerical simulations.

The initial conditions of our numerical simulations correspond to a torus of perfect fluid in hydrostatic equilibrium, orbiting a naked singularity. We followed the theory of such configurations presented in \cite{Abramowicz:1978} for black holes, according to which the perfect fluid in hydrostatic equilibrium will satisfy
\begin{equation}
    \frac{\nabla p}{w} = -\nabla \log{u_t} + \frac{\Omega \nabla l}{1 - \Omega l},\label{eq:Euler_equation}
\end{equation}
where $w$ and $p$ are respectively enthalpy density and pressure of the perfect fluid, $u$ is the four-velocity of the fluid, normalized to $u_\mu u^\mu = -1$ in the signature of the metric (\ref{eq:Reissner_Nordstrom_metric}), while
\begin{align}
\Omega &:= \frac{u^\phi}{u^t}\ , & l := -\frac{u_\phi}{u_t}
\end{align}
can be interpreted as angular frequency in orbital motion of the fluid and the specific angular momentum of a test particle comoving with the fluid.

\begin{figure}[!]
\centering 
\includegraphics[width=\columnwidth]{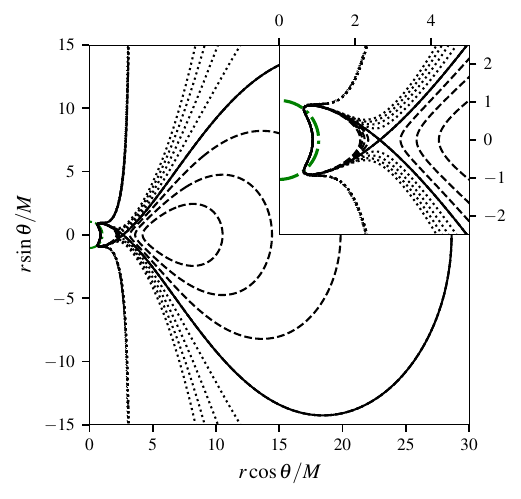}
\caption{Surfaces of constant potential, $W=\mathrm{const}$, at a fixed value of $l_0=3.22$, for Reissner-Nordström NkS with $Q/M = 1.02$. Solid curve: $W = W_\text{cusp}$, i.e. surface of the equilibrium torus with a cusp at $r_\text{cusp}=2.65$. Dashed and dotted curves: surfaces with respectively smaller and higher values of $W$ than $W_\text{cusp}$. The green dashed-dotted line denotes the location of the zero-gravity sphere.}
\label{fig:equlibrium_plot}
\end{figure}

For a specified equation of state $p = p(w)$, for example the polytropic equation of state which we use in our simulations
\begin{equation}
    p = (\Gamma - 1) \varepsilon,
\end{equation}
where $\Gamma$ is the polytropic constant and $\varepsilon$ is the internal energy density of the fluid defined thorough $w = \rho + \varepsilon + p$ with $\rho$ being the mass density,
equation~\eqref{eq:Euler_equation} can be integrated along a~curve to get Boyer's condition:
\begin{equation}
    W := \int \frac{dp}{w} = \log{u_t} - \int \frac{\Omega dl}{1 - \Omega l}.
    \label{eq:Boyers_condition}
\end{equation}
This leads to the conclusion that constant pressure surfaces (isobars) are surfaces of constant $W$ potential.

\begin{figure}[!]
\centering 
\includegraphics[width=\columnwidth]{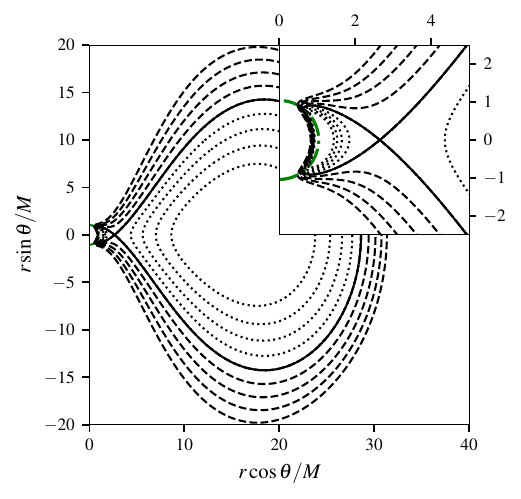}
\caption{Surfaces of constant potential $W$ for Reissner-Nordström NkS with $Q/M = 1.02$. Solid curve: $W = W_\text{cusp}$, surface of an equilibrium torus with the value $l_0=3.22$ in Eq.~\ref{eq:boyer} corresponding to a cusp at $r_\text{cusp}=2.65$. Dashed and dotted curves: surfaces with respectively smaller and larger values of the parameter $l_0$, but the same value of the potential $W=W_\text{cusp}$. The green dashed-dotted line denotes the location of the zero-gravity sphere.
}
\label{fig:disturbed_plot}
\end{figure}

When initializing our simulations, we assumed the simplest case of \mbox{$l = \mathrm{const}=:l_0$} for which the integral on the right hand side of \eqref{eq:Boyers_condition} vanishes and the potential $W$ is given as
\begin{equation}
    \begin{split}
        W = \log{u_t} =& \frac{1}{2} \log{\left(\frac{f(r) r^2 \sin^2{\theta}}{r^2 \sin^2{\theta} - f(r) l_0^2}\right)}\\
        =& \frac{1}{2} \log{\left(\frac{(r^2 - 2Mr + Q^2) \sin^2{\theta}}{r^2 \sin^2{\theta} - (r^2 - 2Mr + Q^2) l_0^2 r^{-2}}\right)}.
    \end{split}
    \label{eq:boyer}
\end{equation}

As a source for the accreting matter we have chosen to use tori with cusps at their inner edge. Since the fluid at the cusp moves freely in Keplerian unstable circular motion, following an infinitesimal perturbation it can plunge in to the other (inner) side of the cusp without any loss of angular momentum. In the simulation, the perturbation is due to small imperfections of our numerical setup. An example of the radial dependence of the effective potential $W$ for one of the values of $l_0$ which we used for numerical simulations is shown in Fig.~\ref{fig:W_example}. For $l_0=3.22$ the potential $W$ has a~local maximum in the equatorial plane at $r = r_\mathrm{cusp} = 2.65 M$, where the cusp is located. The condition for the cusp can be easily computed from \eqref{eq:boyer}:
\begin{equation}
    \left. \frac{\partial W}{\partial r} \right|_{\theta = \pi/2} = 0
\end{equation}
which can be solved exactly for $l_0$, once the position of the cusp is fixed. The value of the potential $W_\text{cusp}$ at the external surface of the torus can be then computed from \eqref{eq:boyer} substituting the correct values of $r_\text{cusp}$ and $l_0$.

Fig.~\ref{fig:equlibrium_plot} presents a cross-section through such a torus in hydrostatic equilibrium in the plane of constant $\phi$ (we assumed that the torus is initially axially symmetric). The solid line is the equipotential surface $W = W_\text{cusp}$. It is worth stressing that the surface has two lobes that form two distinct closed regions (volumes). Both can be external boundaries of fluid bodies in hydrostatic equilibrium filling the two cusped tori. They are connected by a~one dimensional circle (perpendicular to the meridional cross-section presented in the figure), where the equipotential surface self-intersects. In addition, we plotted the equipotential surfaces for a~few smaller values of $W$ as dashed lines. They coincide with isobars for the fluid in hydrostatic equilibrium initially filling the torus. The dotted lines are equipotential surfaces with potential values greater than the value at the surface of the torus.

Fig.~\ref{fig:disturbed_plot} presents how equilibrium configurations depend on the value of $l_0$. The solid curve is the same as in Fig.~\ref{fig:equlibrium_plot}. Dashed curves correspond to equipotential surfaces with the same value of the potential $W=W_\text{cusp}$, but with smaller values of $l_0$. The cusp is present for only one value of $l_0$ for a fixed value of $W=W_\text{cusp}$. As is visible in the plot, for smaller values of $l_0$ the equipotential surface encloses one (simply connected) region and both lobes of the volume enclosed by the self-intersecting surface lie in its interior. On the other hand, for higher values of $l_0$ the equipotential surface has two disconnected parts which may contain two separate equilibrium tori. 

\begin{figure}[!ht]
\centering 
\includegraphics[width=\columnwidth]{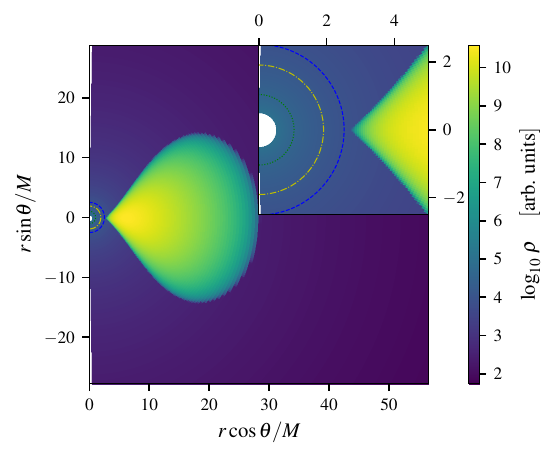}
\includegraphics[width=\columnwidth]{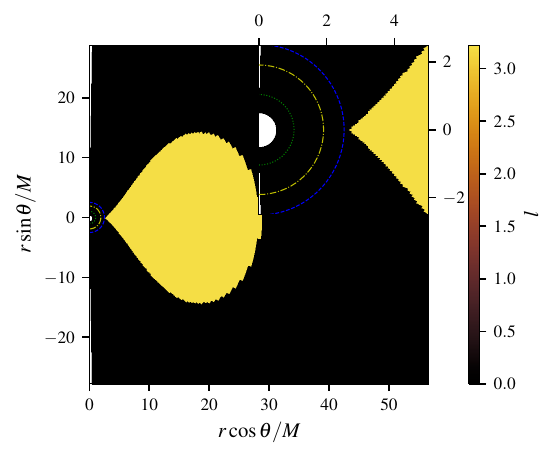}
\caption{Initial conditions ($t=0$) of accretion simulation S02c26 for Reissner-Nordström NkS with $Q/M=1.02$. The cusp of the initial torus is located at $r_\text{cusp}/M=2.65$ (Table~1). {\sl Upper panel:} the logarithm of mass density $\log_{10} \rho$. {\sl Lower panel:} the distribution of the quantity $l =- u_\phi/u_t$, with value $l=l_0$ inside and $l=0$ outside the torus, respectively. The green dotted line indicates the zero-gravity sphere. The dashed blue line represents marginally bound orbits, and the yellow dash-dotted line the unstable photon orbits.}
\label{fig:initial_configuration}
\end{figure}
\section{Simulations\label{sec:sims}}

Simulations were initialized with a torus in hydrostatic equilibrium filling the outer part of an equipotential surface endowed with a cusp, as described in Section \ref{sec:equilibrium_tori}. Outside the torus artificial numerical atmosphere with mass density much smaller then the density in the torus was added. An example of initial conditions for $Q=1.02$ is presented in Fig.~\ref{fig:initial_configuration}.

In order to fully define the problem which we solve numerically one needs to specify, beside initial conditions, also boundary conditions. Our computational domain is a rectangle in internal coordinates of the code $(s, \theta)$, i.e. \mbox{$\{(s, \theta) \colon s \in [s_\text{inner}, s_\text{outer}], \theta \in [\epsilon_\theta, \pi - \epsilon_\theta]\}$}. The small parameter $\epsilon_\theta > 0$ is required to excise the axis, because the $\phi$ coordinate is ill-defined at the poles $\theta=0, \pi$. For the outer edge of the computational domain $s=s_\text{outer}$ we used outflowing boundary conditions which allow induced winds to escape the computational domain. The boundaries close to the polar axis, i.e. $\theta = \epsilon_\theta, \pi - \epsilon_\theta$ were chosen, as usual, as reflective ones. In our studies we assume that the axial symmetry of initial (and boundary) conditions is conserved during the evolution, thus our two dimensional simulations model an axially symmetric three dimensional system.

The most important decision involved the choice of boundary conditions for the inner edge of computation domain, i.e. $s=s_\text{inner}$. We assumed that reflective boundary conditions should be used close to the naked singularity, based on the behaviour of test particles in the region close to the origin of coordinate system. Test particles that come closer than the zero-gravity radius $r_0$ will be repelled due to the repulsive nature of gravity close to the naked singularity, thus the hypothetical test particle that would cross the inner part of the boundary will reenter the computational domain, possibly at some other angular position, but with inverted radial component of the velocity. The selected boundary condition is equivalent to the approximation that particles forming the simulated perfect fluid reenter the computational domain immediately at the same point. 

This choice of inner boundary conditions is much more numerically demanding than the absorbing boundary conditions which are used for simulations of accretion onto black holes. Studies of black holes usually exploit some horizon penetrating coordinate system (coordinate system in which the metric tensor at the event horizon is not singular) with computational domain extending below the horizon. The part of the computational domain below the horizon is causally disconnected from the dynamics modelled outside of it, thus the absorbing boundary conditions are a safe choice for the part of the boundary lying below the horizon, since it will not influence the object of interest---the fluid outside the horizon. In case of simulations of naked singularities, the choice of boundary conditions is important, since any flow with high enough kinetic energy to penetrate the inner part of the boundary will influence the dynamics of the modelled process.

\begin{table}[!h]
\centering
\begin{tabular}{|c|c|c|c|c|}
\hline
\multicolumn{5}{|c|}{{Simulation details}} \\ \hline
Run name & $Q/M$ & $r_\mathrm{cusp}/M$ & $l_0$ & Run time/$M$ \\ \hline
S02c26 & 1.02 & 2.65 & 3.22 & $5\cdot10^4$ \\ \hline
S02c30 & 1.02 & 3.00 & 3.12 & $5\cdot10^4 $  \\ \hline
S07c23 & 1.07 & 2.30 & 3.10 & $5\cdot10^4 $  \\ \hline
S07c25 & 1.07 & 2.50 & 3.04 & $5\cdot10^4 $  \\ \hline
S09c20 & 1.09 & 2.00 & 3.03 & $5\cdot10^4 $  \\ \hline
S09c23 & 1.09 & 2.30 & 3.97 & $5\cdot10^4 $  \\ \hline
\end{tabular}
\caption{Parameters of numerical simulations of accretion onto Reissner-Nordstr\"om naked singularities.}
\label{tab:tableruns}
\end{table}

In order to minimize the influence of (all in all arbitrary) choice of boundary conditions we have chosen the inner edge of computational domain to be equipotential with respect to the outer edge, i.e. a test particle that is at rest at the inner edge, $r=r_\text{inner}$, will reach the outer edge, $r=r_\text{outer}$, with zero velocity. This choice guarantee that the fluid that is initially at rest will penetrate neither the inner nor the outer boundaries unless it first gains kinetic energy.

The price for using a grid-based approach to simulations is that we cannot fully reproduce the physics of perfect-fluid motion close to the boundary since, as was shown by \cite{Vieira:2023cvn}, for any reasonable density of the fluid at the zero-gravity sphere, the density of mass close to the inner part of the boundary should be zero. Grid-based hydrodynamical simulations cannot however work with vacuum, finite difference schemes of modelling the hydrodynamics lead to large relative errors at small densities, and we have to set the mass density in a~vicinity of the inner edge of computational domain to some relatively small, but still finite non-zero value.

Furthermore, the initial stage of relaxation of the initial setup cannot be fully reproduced in our simulations, since the fluid from the initial atmosphere repelled from the neighbourhood of the singularity reaches ultra-relativistic velocities which are above the scope of application of the available numerical conversions from so called conserved to primitive variables (Appendix \ref{sec:numerical_code}). However, this is not restrictive for the object of interest, since the initial low-density atmosphere is already completely artificial.

For uniformity of results, we restrict our simulations to cusped tori. Study of accretion onto RN naked singularities with $Q/M > \sqrt{5}/2$ is beyond the scope of the current paper and will be presented elsewhere. No marginally stable orbits exist for $Q/M > \sqrt{5}/2$, consequently there are no self-intersecting equipotentials, and the equilibrium tori are stable, having no cusps. Simulation of accretion in that regime would require some source of viscosity, such as the magnetorotational instability, which is expected to be the mechanism generating an effective viscosity in ionized accretion disks \citep{Balbus:2003xh}.

Table~\ref{tab:tableruns} gives an overview or the simulations presented in this paper. All runs were for Reissner-Nordstr\"om naked singularities. The simulations differed in the values of two parameters: the spacetime metric was fixed by the value of charge to mass ratio, $Q/M$, and each run initialized with a polytropic torus in hydrostatic equilibrium with a cusp at $r=r_\mathrm{cusp}$. All runs were ended at simulation time $t=5\cdot10^4M$.

\section{Results of numerical simulations\label{sec:numerical_results}}

In the following sections we describe the result of simulations for three choices of the charge to the mass ratio, $Q/M$, which represent three different regimes of stability of circular orbits in the spacetime of the Reissner-Nordström metric, depicted in Fig.~\ref{fig:stability_plot}. The first one, $Q/M = 1.02$, corresponds to the case when there are no circular timelike geodesics in some part of the spacetime (between photon orbits with radii $r_{\gamma\pm}$); there are marginally bound orbits, thus circular orbits between them have energies greater than rest mass of the particle at infinite distance; and all circular orbits with radius larger than $r_\gamma$ but smaller than $r_{\text{OMSCO}}$ are unstable---in short, $r_\gamma$, $r_{\text{mb}}$ and $r_{\text{ms}}$ all exist. For the second case, $Q/M=1.07$, circular timelike orbits exist for all radii up to the zero-gravity radius $r_0$ (there is no photon orbit), but some of them are unbound and some of them unstable ($r_{\text{mb}}$ and $r_{\text{ms}}$ still exist). The last case $Q/M=1.09$ is the metric for which circular timelike orbits exist with radii up to zero-gravity radius $r_0$ and all are bound, but some of them are unstable ($r_{\text{ms}}$ alone is present). As described in Section~\ref{sec:stability_analysis}, when $Q/M > \sqrt{5}/2$ all circular orbits with radii greater than the zero-gravity radius $r_0$ are stable, so no cusped equilibrium tori exist in this regime (Section~\ref{sec:equilibrium_tori}). 

\begin{figure}[!ht]
\centering 
\includegraphics[width=\columnwidth]{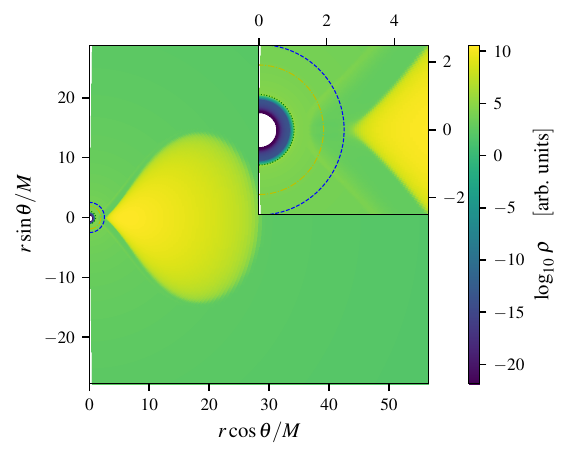}
\caption{The logarithm of rest mass density, $\log_{10} \rho$, in accretion onto a Reissner-Nordström NkS ($Q/M=1.02$). The data are a~snapshot of simulation S02c26 results at time $t=10\, t_g$.}
\label{fig:early_levitating atmosphere}
\end{figure}

\subsection{Case study: $Q/M=1.02$\label{sec:Q=1.02}}
The initial conditions for one of our simulations with $Q/M=1.02$ is presented in Fig.~\ref{fig:initial_configuration}. In this example, simulation S02c26, we used an equilibrium torus with a cusp located at the radius $r_\text{cusp}/M = 2.65$ ($l_0 = 3.22$) which gives a torus of reasonable size for numerical simulations.

Fig.~\ref{fig:early_levitating atmosphere} presents configuration of the fluid at time $t=10\ t_g$. A levitating atmosphere \citep{Vieira:2021nzs,Vieira:2023cvn} is visible around the zero-gravity sphere (dotted green line); it was formed from matter belonging initially to the artificial atmosphere, which was repelled from the vicinity of the naked singularity. 

\begin{figure}[!ht]
\centering 
\includegraphics[width=\columnwidth]{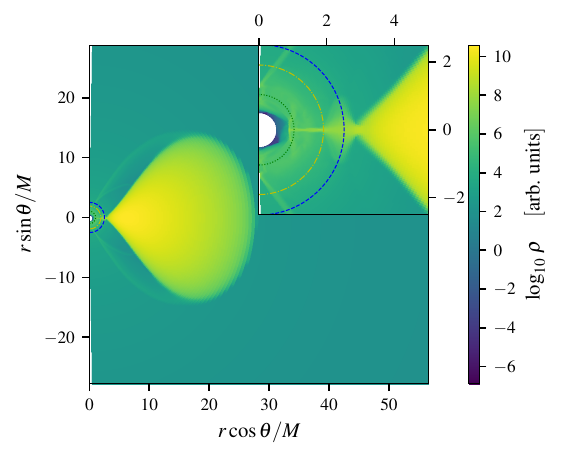}\\
\includegraphics[width=\columnwidth]{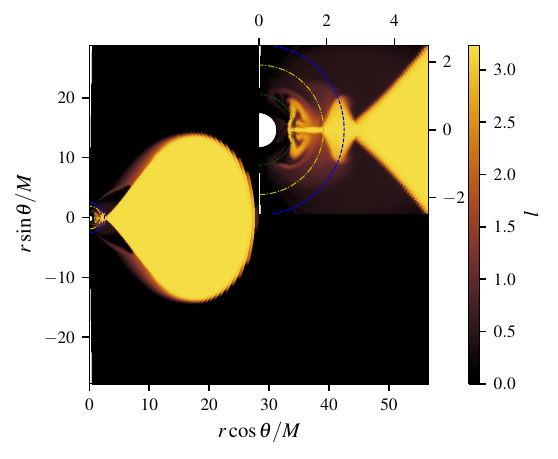}
\caption{A snapshot of accretion at time $t=160\, t_g$, Reissner-Nordström NkS with $Q/M=1.02$, simulation S02c26. {\sl Upper panel:} logarithm of rest mass density, $\log_{10} \rho$. {\sl Lower panel:} the quantity $l$. The circular dotted green line corresponds to the zero-gravity sphere. The yellow dash-dotted line indicates unstable photon orbits, the dashed blue line--marginally bound orbits.}
\label{fig:early_accretion}
\end{figure}
The region below the levitating atmosphere should contain vacuum, however numerical methods for hydrodynamics used in our studies cannot properly evolve such a configuration and this region is instead filled with a fluid of density set by a numerical floor, as described in Appendix \ref{sec:numerical_code}. 
We believe that this imperfection of the numerical scheme has not affected our final findings about the nature of the quasi steady-state accretion, since the density of fluid forming the final configuration is many orders of magnitude greater than the minimal value of the mass density necessitated by the numerical scheme. Our belief is supported by the results of the simulations that we run varying some of the numerical floors (especially the one limiting the mass density $\rho$) for the described case study. We have not observed qualitative changes in the behaviour of the accretion even though the investigated values of numerical floors ranged over dozens of orders of magnitude. On the other hand we observed that certain numerical floors (especially the one limiting the relativistic Lorentz factor of the fluid velocity) are crucial for simulating the relaxation of the initial conditions toward steady-state accretion and raising them leads to a breakdown of the numerical scheme for the conversion from conserved variables to primitive ones.

\begin{figure}[!ht]
\centering 
\includegraphics[width=\columnwidth]{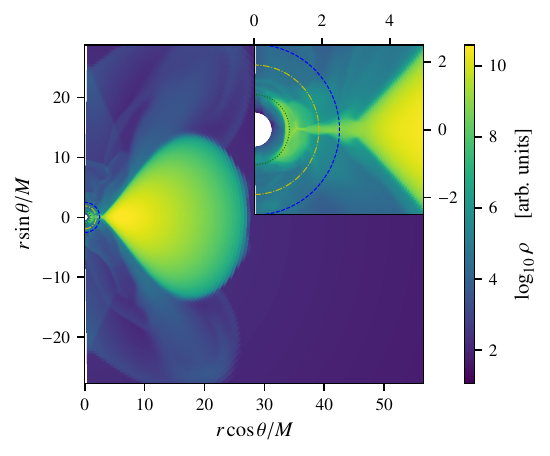}
\caption{Accretion onto the $Q/M=1.02$, Reissner-Nordström NkS. The data are a snapshot of simulation S02c26 results at time $t=310\, t_g$. The plot presents the logarithm of rest mass density, $\log_{10} \rho$. The circular lines have the same meaning as in Fig.~\ref{fig:early_accretion}.}
\label{fig:early_accumulation}
\end{figure}

During the next $150\,t_g$ the material starts to accrete from the torus toward the zero-gravity sphere, as may be seen in Fig.~\ref{fig:early_accretion}. It is worth stressing that material slightly penetrates the zero-gravity sphere, reaching the inner edge of the inner branch of the cusped equipotential surface presented in Fig.~\ref{fig:equlibrium_plot}. From the lower panel of Fig.~\ref{fig:early_accretion} we deduce that initially the matter does not loose much angular momentum, $l$.
After the next $150\,t_g$, the fluid starts to gather around the zero-gravity sphere as is visible in Fig.~\ref{fig:early_accumulation}. Moreover, some material is ejected and forms a~wind around the torus.

\begin{figure}[!ht]
\centering 
\includegraphics[width=\columnwidth]{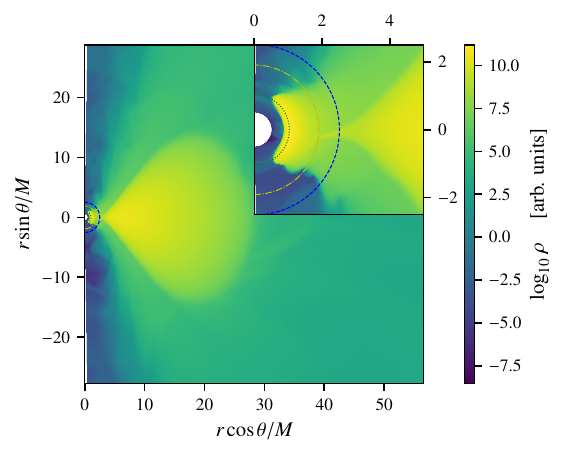}\\
\includegraphics[width=\columnwidth]{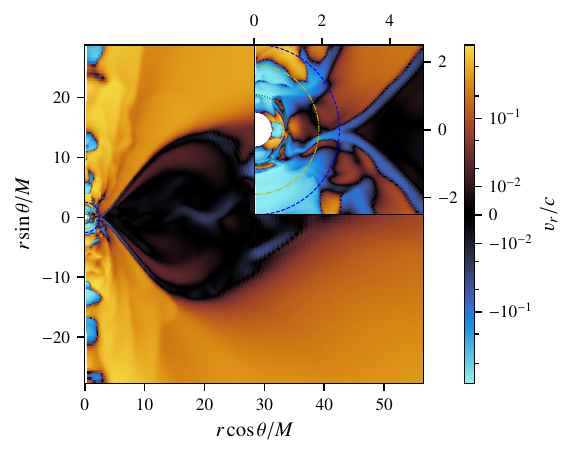}
\caption{Steady state of accretion onto the $Q/M=1.02$, Reissner-Nordström NkS. A snapshot of simulation S02c26 results at time $t=10^4 t_g$ is shown. {\sl Upper panel:} logarithm of rest mass density, $\log_{10} \rho$. {\sl Lower panel:} radial component of the fluid velocity, $\varv_r=-u^r/u_t$, in units of the speed of light.
The circular lines have the same meaning as in Fig.~\ref{fig:early_accretion}.}
\label{fig:q=1.02_rcusp=2.65_steady_state}
\end{figure}

After some time ($\sim 10^3\,t_g$), the described initial transients decay away and the flow stabilizes. An example of this later stage of evolution is presented in the snapshot of Fig.~\ref{fig:q=1.02_rcusp=2.65_steady_state}. In the torus the flow of matter toward the naked singularity occurs at its surface, as is visible in the bottom panel of Fig.~\ref{fig:q=1.02_rcusp=2.65_steady_state}, which presents the radial component $\varv_r=-u^r/u_t$ of the velocity of the perfect fluid. The plasma leaves the torus at its cusp, and forms a geometrically thin accretion stream which is deflected from the equatorial plane. The orientation of the deflection changes through the simulation. This suggests that flow along the equatorial plane is probably unstable and a spontaneous symmetry breaking occurs due to imperfections of the numerical scheme amplified by the chaotic character of flow equations.

As one can expect from the considerations in Section \ref{sec:stability_analysis}, owing to the repulsive nature of the gravitational field in the close neighbourhood of the singularity the accreted matter never gets much closer to the singularity than the zero-gravity sphere and, in fact, it accumulates on both sides of it, as is visible in the upper panel of Fig.~\ref{fig:q=1.02_rcusp=2.65_steady_state}, which shows the distribution of the rest mass density $\rho$. The accumulated matter forms another torus around the naked singularity which we will call the inner torus, to distinguish it from the one set up during initialization as a reservoir of accreting matter, we will denote the latter one as the outer torus from now on.

\begin{figure}[!]
\centering 
\includegraphics[width=\columnwidth]{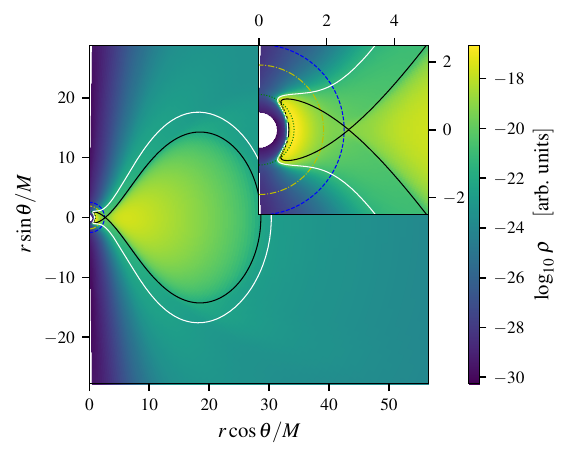}
\caption{Steady state of accretion onto Reissner-Nordström NkS, $Q/M=1.02$. Simulation S02c26. The logarithm of rest mass density $\log_{10} \rho$ averaged over simulation time from $t=10^4 t_g$ to $t=5 \times 10^4 t_g$ is shown. Black lines correspond to equipotential surface of initial torus. White lines denote surfaces of constant potential value equal to that of the initial torus surface, but with a value of the parameter $l_0$ equal to an average of $l$ over the dense region close to the naked singularity. The circular lines have the same meaning as in Fig.~\ref{fig:early_accretion}.}
\label{fig:q=1.02_rcusp=2.65_steady_state_rho}
\end{figure} 

Time averaged data from our simulations are presented in Fig.~\ref{fig:q=1.02_rcusp=2.65_steady_state_rho}. The black lines represent the equipotential curve corresponding to the surface of the initial torus. As was discussed in Section \ref{sec:equilibrium_tori}, the presented surface (a curve in the meridional plane shown in the figures) encloses two compact regions which are connected by the circle of its self-intersection, where the two cusps of the outer and the inner regions osculate. The initial torus used in our simulations fills the outer region, however both regions can host equilibrium tori. One may naively expect that the accretion will proceed due to simple flow of material from the outer region to the inner region through the cusps. Even though this is the case initially, our simulations show that later on the process is not necessary so simple. As is visible in the upper panel of Fig.~\ref{fig:q=1.02_rcusp=2.65_steady_state_rho} the inner torus does not fit inside the equipotential surface enclosing the initial torus, and some matter gathers outside that surface.

\begin{figure}[!]
\centering 
\includegraphics[width=\columnwidth]{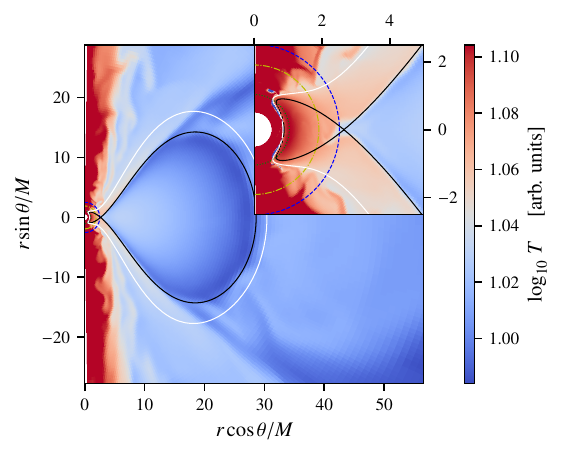}\\
\includegraphics[width=\columnwidth]{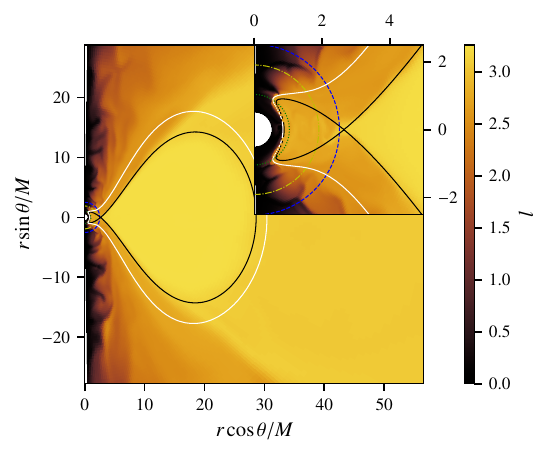}
\caption{Steady state of accretion onto the $Q/M=1.02$ Reissner-Nordström NkS. The data are a snapshot of simulation S02c26 results at time $t=5 \times 10^4 t_g$. {\sl Upper panel:} logarithm of the temperature of fluid $\log_{10} T$. {\sl Lower panel:} the quantity $l$. Black lines correspond to the equipotential surface of initial torus. White lines denote surfaces of constant potential value equal to that of the initial torus surface, but with a value of the parameter $l_0$ equal to an average of $l$ over the dense region close to the naked singularity. The circular lines have the same meaning as in Fig.~\ref{fig:early_accretion}.}
\label{fig:q=1.02_rcusp=2.65_steady_state_temperature}
\end{figure}

From inspection of the lower panel of Fig.~\ref{fig:q=1.02_rcusp=2.65_steady_state_temperature} one may deduce that the fluid in the inner torus has a nearly constant value of $l$, which is however smaller than the value in the initial torus. Furthermore, one can see in the inset of the lower panel of Fig.~\ref{fig:q=1.02_rcusp=2.65_steady_state_temperature} that the material around the surface of the inner torus has an even lower value of $l$, since the fluid between the white and black line has a darker shade in our color scheme than the accretion stream of plasma leaking from the outer torus.

The cusp of the outer torus is necessarily within the marginally stable orbit, here $r_\mathrm{cusp} <r_\mathrm{OMSCO}$, so the infalling fluid cannot find a circular orbit corresponding to its value of $l$ until it reaches the inner marginally stable orbit at $r_\mathrm{EMSCO}$, having first traversed the region forbidden to circular orbits between the (inner, stable and outer, unstable) photon orbits ($Q/M=1.02$, Fig.~\ref{fig:stability_plot}). A more detailed investigation reveals that in the simulation the fluid in the inner torus has a lower value of $l$ than the initial torus. 

We estimated the $l$ value of the material gathered in the inner torus by applying a low rest-mass density $\rho$ mask and averaging over the unmasked region below the radius of the location of the cusp of the initial torus. The equipotential curve corresponding to the same value of the potential as the initial torus, but with $l$ of the inner torus determined from simulations, is plotted in Fig.~\ref{fig:q=1.02_rcusp=2.65_steady_state_rho} as a~white line. It presents remarkable compatibility with the surface of the inner torus close to the naked singularity. The region between the two equipotential surfaces (white and black curves in the inset) is filled with fluid in motion.

We expect that the deflection of accreting matter from the equatorial plane is caused by the pressure of the matter already accumulated closer to the naked singularity. Had we started our simulations with both parts of the equilibrium curves filled with the fluid in equilibrium (not only outer torus but also the inner one) the accreting matter would move above the surface, i.e. along the equipotential curves enclosing the fluid which diverge from the equatorial plane at the cusp.

The upper panel of Fig.~\ref{fig:q=1.02_rcusp=2.65_steady_state_temperature} presents the logarithm of the temperature of the fluid, which we present as an easy to visualize proxy for the internal energy of the fluid. A brief inspection of the plot reveals that the inner torus is hotter (has larger internal energy), than the initial one. Furthermore, we observe that the stream of matter leaking from the outer torus is heating up when moving toward the naked singularity. This would explain why $l = - u_\phi / u_t$ decreases during the accretion process. The proper expression for the conserved angular momentum density associated with $\partial_\phi$ Killing vector is $j := T^t_\phi = w u^t u_\phi$ which is proportional to the enthalpy $w$. Recalling that $w = \rho + \varepsilon + p$ we notice that the enthalpy grows with incensing temperature $T$, since both internal energy density $\varepsilon$ and pressure $p$ are increasing functions of $T$. For $j$ to remain constant while the fluid is being heated, $|u^t u_\phi|$ needs to decrease, so $|l| = |- u_\phi / u_t|$ will necessarily decrease as well.

\begin{figure}[!]
\centering 
\includegraphics[width=\columnwidth]{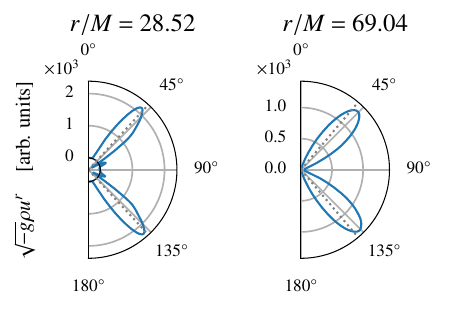}\\
\includegraphics[width=\columnwidth]{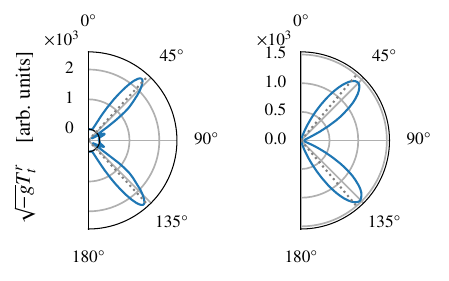}\\
\includegraphics[width=\columnwidth]{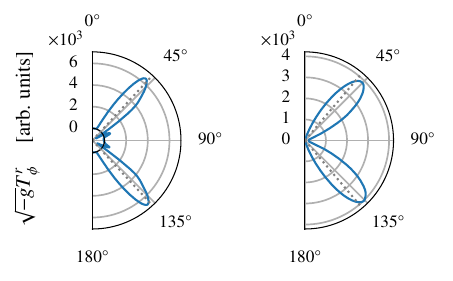}
\caption{Angular distribution of fluxes of rest mass $\sqrt{-g} \rho u^r$ {\sl(top panel)}, energy $\sqrt{-g} T^r_t$ {\sl(middle panel)}, and angular momentum $\sqrt{-g} T^r_\phi$ {\sl(bottom panel)} through the spherical surfaces with radii $r/M = 28.52$ (outer edge of the initial torus) and $r/M = 69.04$ (a few computational cells from outer boundary) for the Reissner-Nordström $Q/M=1.02$ simulation S02c26. The data are the averages of simulation results from time $t=10^4 t_g$ up to $t=5 \times 10^4 t_g$. Grey dashed lines delimit the (imaginary) shadow of the torus cast by the cusp.}
\label{fig:q=1.02_rcusp=2.65_fluxes}
\end{figure}
The angular distribution of the radial component of fluxes of rest mass density, energy, and angular momentum, respectively $\sqrt{-g}\, \rho u^r$, $\sqrt{-g}\,T^r_t$, $\sqrt{-g}\,T^r_\phi$, are presented in Fig.~\ref{fig:q=1.02_rcusp=2.65_fluxes}. The left column of Fig.~\ref{fig:q=1.02_rcusp=2.65_fluxes} presents the fluxes at $r=28.52$, which is the maximal radial extent of the initial torus, while the right column shows values computed at $r=69.04$, which is close to the outer boundary of the computational domain, but separated from it by a few computational cells in order to avoid boundary effects on the presented data. The three presented fluxes are highly correlated with each other, showing a very similar angular distribution. They are collimated at an angle of around $45^\circ$ from the equatorial plane, with the polar angle of maximal flux somewhat lower closer to the naked singularity. The grey dotted lines in Fig.~\ref{fig:q=1.02_rcusp=2.65_fluxes} marks the edge of the shadow of the initial torus as seen from the location of the cusp, i.e. a line passing through the location of the cusp and tangent to the surface of the initial torus at a point $(r, \theta)$, with the value $\theta$ presented on the polar plots. The left column of Fig.~\ref{fig:q=1.02_rcusp=2.65_fluxes} shows that the outflow of matter proceeds along the surface of the outer torus. At higher distances, judging from the right column, it is slightly deflected toward the equatorial plane and diffused somewhat.

\begin{figure}[!]
\centering 
\includegraphics[width=\columnwidth]{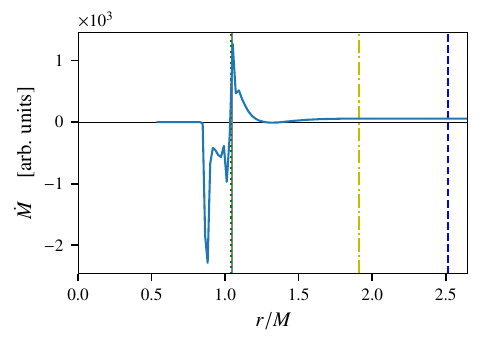}\\
\includegraphics[width=\columnwidth]{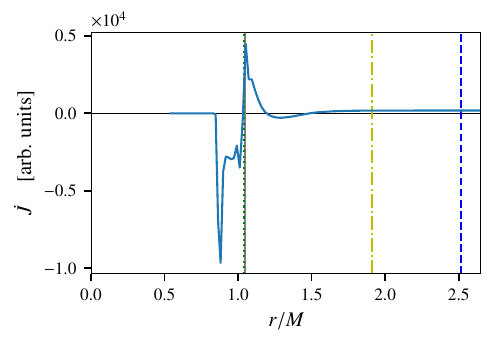}
\caption{The rest mass accretion rate $\dot{M}$ {\sl(upper panel)} and the angular momentum accretion rate $\dot{J}$ {\sl(lower panel)} as a function of the coordinate distance $r/M$ from the naked singularity in steady state of the simulated accretion onto the $Q/M=1.02$ Reissner-Nordström NkS. The data are averages of simulation S02c26 results over time $t=10^4 t_g$ to $t=5 \times 10^4 t_g$. The green dotted line indicates the radius of the zero-gravity sphere, the yellow dash-dotted line the radius of the unstable photon orbit, and the dashed blue line the radius of marginally bound orbits. The grey solid line marks the location of the maximum of the pressure in the equatorial plane.}
\label{fig:q=1.02_rcusp=2.65_steady_state_accretion_rate}
\end{figure}
In Fig.~\ref{fig:q=1.02_rcusp=2.65_steady_state_accretion_rate} we present the accretion rates $\dot{M}$ and $\dot{J}$ of the rest mass and the angular momentum respectively as a function of the coordinate distance $r$ from the naked singularity, averaged over time from $t=10^4 t_g$ to $t=5 \times 10^4 t_g$. One can conclude from Fig.~\ref{fig:q=1.02_rcusp=2.65_steady_state_accretion_rate} that the accretion rates of both the rest mass and the angular momentum are highly correlated, as was also visible in Fig.~\ref{fig:q=1.02_rcusp=2.65_fluxes}. The grey solid vertical lines on plots in Fig.~\ref{fig:q=1.02_rcusp=2.65_steady_state_accretion_rate} mark the location of the pressure maximum in the equatorial plane, i.e. the ``center'' of the inner torus. As can be seen from Fig.~\ref{fig:q=1.02_rcusp=2.65_steady_state_accretion_rate}, the roots of accretion rate are nearly coincident with the pressure maximum; the accreted matter gathers around this location, filling in the inner torus.

\begin{figure}[!]
\centering 
\includegraphics[width=\columnwidth]{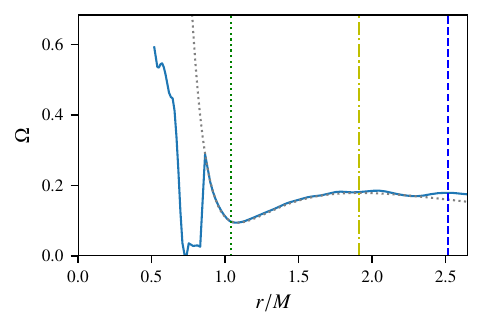}
\caption{Orbital angular frequency, $\Omega$, in the steady state of accretion onto the Reissner-Nordström $Q/M=1.02$ naked singularity as a function of the coordinate distance $r/M$, at time $t=5 \times 10^4 t_g$ of simulation S02c26. The dotted grey line on the main plot is $\Omega$ for a constant value of the quantity $l$ estimated from the conditions in the inner torus. The vertical lines have the same meaning as in Fig.~\ref{fig:q=1.02_rcusp=2.65_steady_state_accretion_rate}.}
\label{fig:q=1.02_rcusp=2.65_steady_state_omega}
\end{figure}

Fig.~\ref{fig:q=1.02_rcusp=2.65_steady_state_omega} shows the profile in the equatorial plane of the orbital frequency $\Omega = u^\phi / u^t$ of the fluid. The grey dotted curve is the expected profile for the fluid with constant $l$, with its value fixed as the average over the inner torus. The fact that this predicted profile agrees in the span of the inner torus with $\Omega$ extracted directly from the simulation data support our hypothesis that the $l$ parameter is constant in the inner torus and that our procedure of estimating its value using the high density mask gives correct results. 

\begin{figure}[!]
\centering 
\includegraphics[width=\columnwidth]{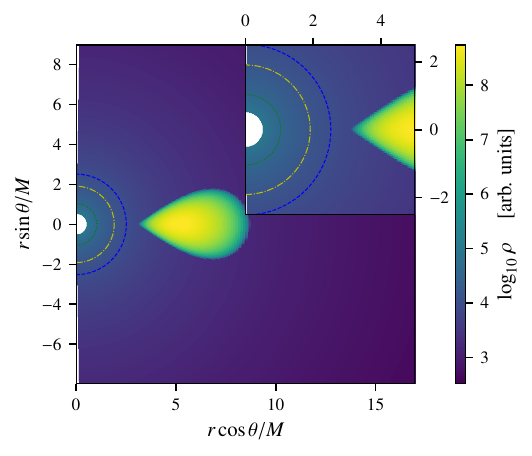}
\caption{Logarithm of rest-mass density $\log_{10} \rho$ in the initial conditions of simulation for the Reissner-Nordström NkS with $Q=1.02$, and the cusp of the initial torus at $r_\text{cusp}/M=3.00$ (simulation S02c30). The green dotted line correspond to zero-gravity sphere and yellow dash-dotted line is the location of the unstable photon orbit. The dashed blue line represents the radius of marginally bound orbits.}
\label{fig:q=1.02_rcusp=3._initial_state}
\end{figure}

We also performed simulations for the torus with the cusp located a little farther from the naked singularity, at \mbox{$r_\text{cusp} / M = 3.00$} (simulation S02c30). The initial conditions for the rest mass density, $\rho$, is presented in Fig.~\ref{fig:q=1.02_rcusp=3._initial_state}. The torus for this setup is smaller and thinner in the direction perpendicular to the equatorial plane (symmetry plane) than the one in the previously considered case.

\begin{figure}[!]
\centering 
\includegraphics[width=\columnwidth]{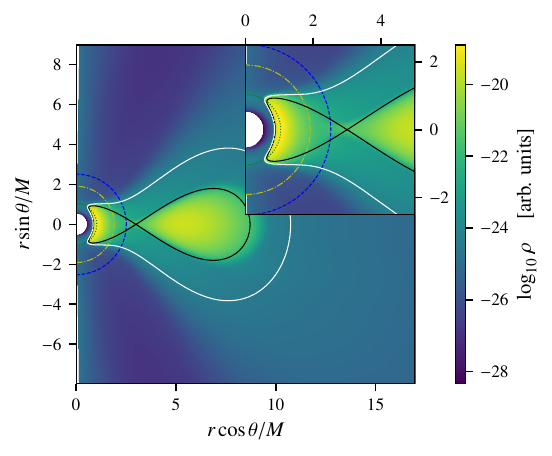}
\caption{Steady-state accretion onto Reissner-Nordström NkS with $Q/M=1.02$. The data are averages of simulation S02c30 results over time from $t=10^4 t_g$ to $t=5 \times 10^4 t_g$. The plot presents the logarithm of rest-mass density $\log_{10} \rho$. Black lines correspond to equipotential surface of initial torus. White lines correspond to  equipotential surfaces with potential value equal to that of the initial torus surface, but with the parameter $l_0$ computed as an average of $l$ over the highly dense region close to naked singularity.}
\label{fig:q=1.02_rcusp=3._steady_state_rho}
\end{figure}

\begin{figure}[!]
\centering 
\includegraphics[width=\columnwidth]{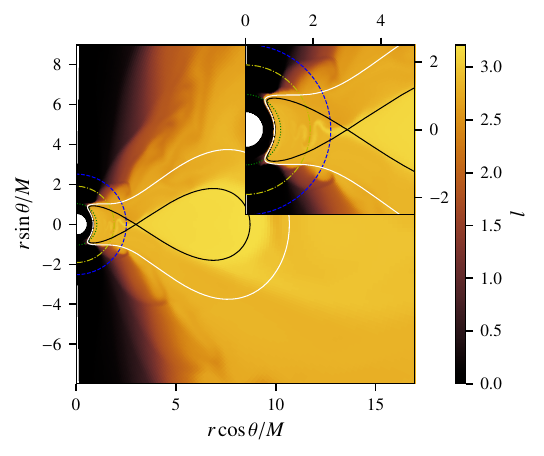}\\
\includegraphics[width=\columnwidth]{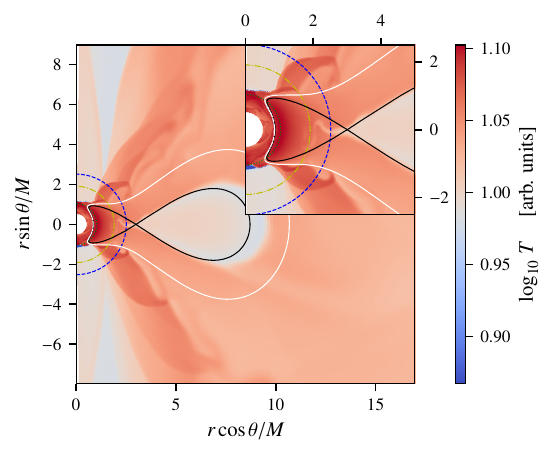}\\
\includegraphics[width=\columnwidth]{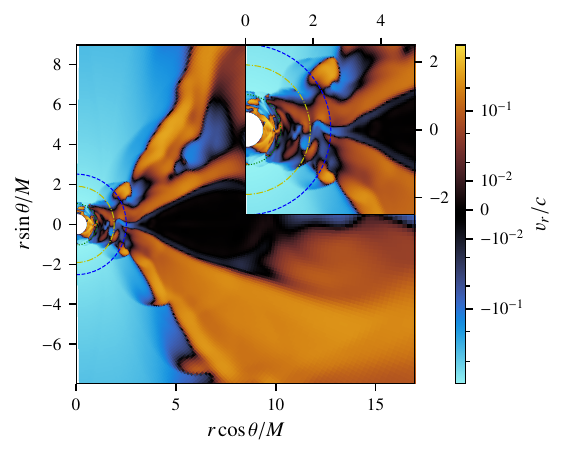}
\caption{Steady state of accretion onto a Reissner-Nordström NkS with $Q/M=1.02$. The data are a snapshot of simulation S02c30 results at time $t=5 \times 10^4 t_g$.
{\sl Upper panel:} value of $l = -u_\phi / u_t$. 
{\sl Middle panel:} logarithm of the temperature, $\log_{10} T$. 
{\sl Lower panel:} radial component $\varv_r=-u^r/u_t$ of the velocity of the perfect fluid. 
The meaning of the lines is the same as in Fig.~\ref{fig:q=1.02_rcusp=3._steady_state_rho}.}
\label{fig:q=1.02_rcusp=3._steady_state_equilibrium_curves}
\end{figure}

The steady-state rest mass distribution $\rho$ for this case averaged over time from $t=10^4\ t_g$ up to $t=5 \times 10^4\ t_g$ is depicted in Fig.~\ref{fig:q=1.02_rcusp=3._steady_state_rho}. The data from snapshot at $t=5 \times 10^4\ t_g$ are presented in Fig.~\ref{fig:q=1.02_rcusp=3._steady_state_equilibrium_curves}. The presented equilibrium curves generated in the same procedure as previously seem to approximately fit the shape of the inner torus and the outflows.

 The distribution of temperature, depicted in the middle panel of Fig.~\ref{fig:q=1.02_rcusp=3._steady_state_equilibrium_curves} is more complex than in the previous case, and this can shed some light on the possible discrepancy between the volume occupied by the fluid and the predicted shape of the its equipotential surface. The high temperature and high mass density region of the inner torus is enclosed by the proposed equilibrium curve. 

The lower panel of Fig.~\ref{fig:q=1.02_rcusp=3._steady_state_equilibrium_curves} shows the radial component $\varv_r$ of the velocity. The outflowing low-density, hot wind is visible, stretching from the surface of the outer torus up to moderate inclination angles. In polar regions, we observe the inflow of the very low density matter coming from the artificial atmosphere. 

\begin{figure}[!]
\centering 
\includegraphics[width=\columnwidth]{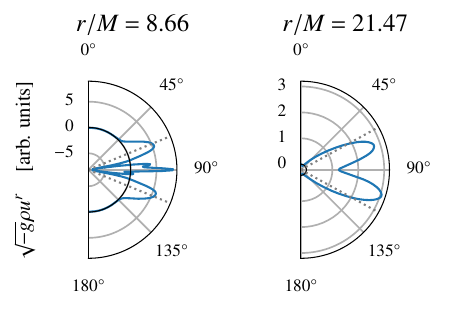}
\caption{Angular distribution of the radial flux of rest mass, $\sqrt{-g} \rho u^r$, through the spherical surfaces with radii $r/M =8.66$ (outer edge of the initial torus) and $r/M = 21.47$ (a few computational cells from outer boundary). The data are time averages of simulation S02c30 results, from $t=10^4 t_g$ to $t=5 \times 10^4 t_g$.}
\label{fig:q=1.02_rcusp=3._fluxes}
\end{figure}

The angular distribution of radial flux of rest mass $\sqrt{-g}\, \rho u^r$, averaged over time from $t=10^4\ t_g$ up to $t=5 \times 10^4\ t_g$ is presented in Fig.~\ref{fig:q=1.02_rcusp=3._fluxes}. The angular distance of maxima of fluxes from the equatorial plane is smaller than in the previous case of $r_\text{cusp}/M=2.65$. This is an expected difference, under the hypothesis that the geometry of the outflows is dictated by the opening angle of the torus as seen from the location of the cusp, since the smaller torus for $r_\text{cusp}/M=3.00$ is thinner, and has a smaller opening angle.

\begin{figure}[!]
\centering 
\includegraphics[width=\columnwidth]{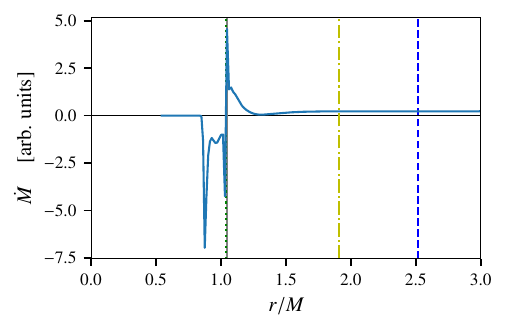}
\caption{The rest mass accretion rate $\dot{M}(r)$ in steady state of accretion onto $Q/M=1.02$ Reissner-Nordström naked singularity averaged over time $t=10^4 t_g$ to $t=5 \times 10^4 t_g$ in simulation S02c30. The vertical lines have the same meaning as in Fig.~\ref{fig:q=1.02_rcusp=2.65_steady_state_accretion_rate}.}
\label{fig:q=1.02_rcusp=3._steady_state_accretion_rate}
\end{figure}

The rest mass accretion rate $\dot{M}$ averaged over time from \mbox{$t=10^4\ t_g$} up to $t=5 \times 10^4\ t_g$ is presented for simulation S02c30 ($Q/M=1.02$ and $r_\text{cusp}/M=3.00$) in Fig.~\ref{fig:q=1.02_rcusp=3._steady_state_accretion_rate} as a~function of the coordinate distance $r$ from the naked singularity. The accretion rate has a root at the position near the location of the center of the inner torus, i.e. the location of the maximum of the pressure, denoted by the grey vertical line in Fig.~\ref{fig:q=1.02_rcusp=3._steady_state_accretion_rate}. The same is true for $\dot J$ (Fig.~\ref{fig:q=1.09_rcusp=2.3_steady_state_accretion_rate}).

\begin{figure}[!]
\centering 
\includegraphics[width=\columnwidth]{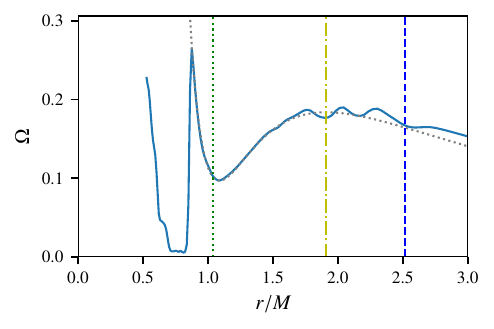}
\caption{The orbital angular frequency, $\Omega(r)$, in steady state of simulation S02c30 for the Reissner-Nordström $Q/M=1.02$ naked singularity at time $t=5 \times 10^4 t_g$. The dotted grey curve is the frequency $\Omega$ for a constant value of the $l$ parameter estimated from the conditions in the inner torus. The vertical lines have the same meaning as in Fig.~\ref{fig:q=1.02_rcusp=2.65_steady_state_accretion_rate}.}
\label{fig:q=1.02_rcusp=3._steady_state_omega}
\end{figure}

The frequency $\Omega$ of the orbital motion of the prefect fluid in the equatorial plane is presented in Fig.~\ref{fig:q=1.02_rcusp=3._steady_state_omega} for time \mbox{$t=5 \times 10^4\ t_g$}. The grey dotted curve, tracking $\Omega(r)$ over the maxima and minimum of this plot at $0.8<r/M<2.5$, corresponds to $l=-u_\phi/u_t=\mathrm{const}$, with the value of $l$ calculated as an average over the inner torus, as discussed previously. The overlap of this curve with the blue solid one representing the simulation data support our hypothesis of constant value of the $l$ parameter inside the inner torus.

\subsection{Case study: $Q/M=1.07$}

The case of $Q/M=1.07$ corresponds to the situation when there are no photon orbits in the spacetime described by the Reissner-Nordström metric, but some unstable orbits have energies higher than a particle at rest at infinite distance from the naked singularity, i.e. there exist marginally bound circular orbits. The aim of studying the case $Q/M=1.07$ is to understand the influence on the process of accretion of (the lack of) forbidden radial intervals for test-particle circular orbits (in the region between the photon orbits) that were present in the previous cases.

\begin{figure}[!]
\centering 
\includegraphics[width=\columnwidth]{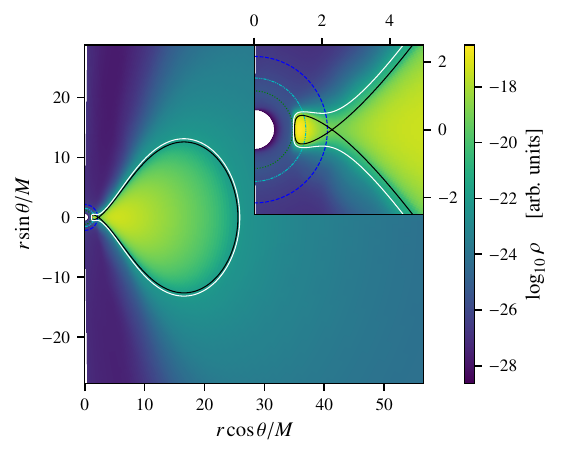}
\caption{Steady state of accretion onto the naked singularity of the Reissner-Nordström spacetime with $Q/M=1.07$. The plot presents the logarithm of rest mass density $\log_{10} \rho$ averaged over time from $t=10^4 t_g$ to $t=5 \times 10^4 t_g$. Black lines correspond to the equipotential surface of the initial torus. White line corresponds to the surface of constant potential with value equal to that of the initial torus surface, with the parameter $l_0$ computed as an average of $l$ over the highly dense region close to naked singularity. The green dotted line corresponds to zero-gravity sphere and the cyan double-dotted dashed line is the location of maximum of Keplerian frequency $\Omega_\mathrm{K}$. The dashed blue line represents the radius of marginally bound orbits.}
\label{fig:q=1.07_rcusp=2.3_steady_state_rho}
\end{figure} 

\begin{figure}[!]
\centering 
\includegraphics[width=\columnwidth]{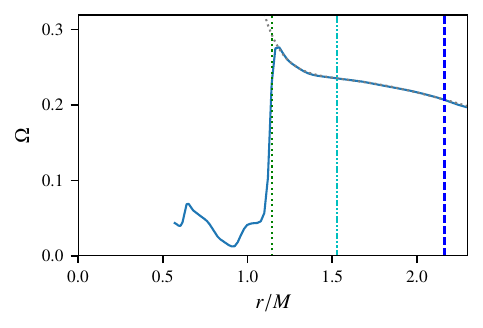}
\caption{The orbital angular frequency (solid curve) in the steady state of accretion onto the $Q/M=1.07$ Reissner-Nordström naked singularity, as a function of the coordinate distance $r/M$, at time $t=5 \times 10^4 t_g$. The cusp of the initial torus is located at $r_\text{cusp}/M=2.30$. The radius of the zero-gravity sphere is indicated by a green dotted line, of maximum Keplerian frequency by the cyan double dotted dashed line, and of marginally bound orbits by the dashed blue line. The dotted grey curve (overlapping the solid curve in the right part of the plot) is the frequency $\Omega$ corresponding to a constant value of the parameter $l=\mathrm{const}$ estimated for the fluid in the inner torus.}
\label{fig:q=1.07_rcusp=2.3_steady_state_omega}
\end{figure} 

\begin{figure}[!]
\centering 
\includegraphics[width=\columnwidth]{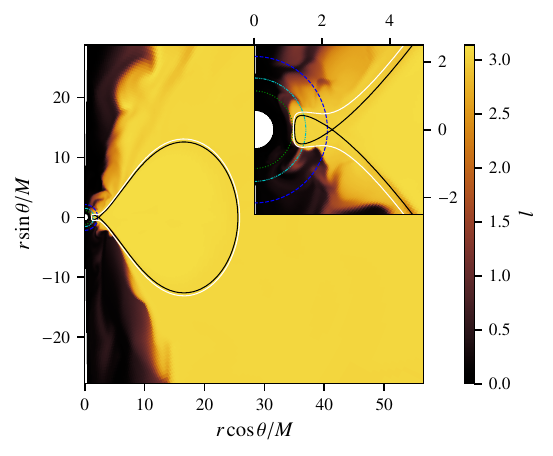}\\
\includegraphics[width=\columnwidth]{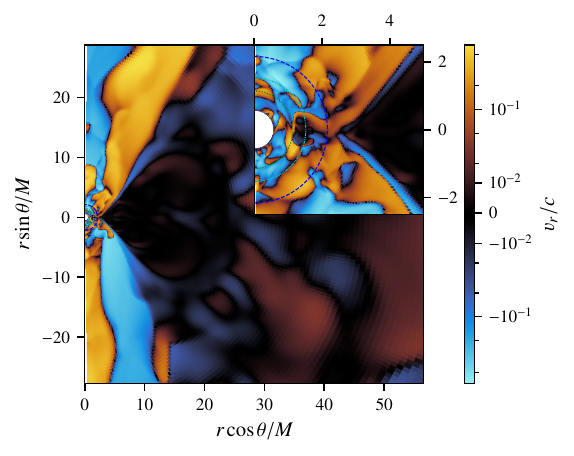}
\caption{Steady-state accretion onto the Reissner-Nordström $Q/M=1.07$ naked singularity. The data are a snapshot of the results at time $t=5 \times 10^4 t_g$.
{\sl Upper panel:} value of $l = -u_\phi / u_t$. 
{\sl Lower panel:} radial component $\varv_r=-u^r/u_t$ of velocity. 
The meaning of the lines is the same as in Fig.~\ref{fig:q=1.07_rcusp=2.3_steady_state_rho}.}
\label{fig:q=1.07_rcusp=2.3_steady_state_equilibrium_curves}
\end{figure} 

The outer cusped equilibrium torus for $Q/M=1.07$ and $r_\text{cusp}/M=2.30$ is similar in size and shape to the equilibrium torus for $Q/M=1.02$ with $r_\text{cusp}/M=2.65$. The inner lobe of the self-intersecting equipotential surface for $Q/M=1.07$ and $r_\text{cusp}/M=2.30$ is however smaller than the one for $Q/M=1.02$.

The rest mass density distribution averaged over time from $t=10^4 t_g$ up to $t=5 \times 10^4 t_g$ is presented in Fig.~\ref{fig:q=1.07_rcusp=2.3_steady_state_rho}. The equilibrium curve depicted as white line in Fig.~\ref{fig:q=1.07_rcusp=2.3_steady_state_rho}, constructed from the value of the $l$ parameter, as averaged over the inner torus, encloses the high density region close to the naked singularity. Strong evidence for a uniform value of the $l$ parameter inside the inner torus can be found in Fig.~\ref{fig:q=1.07_rcusp=2.3_steady_state_omega}, where the orbital angular frequency $\Omega$ is plotted as a~function of the coordinate distance $r$ from the naked singularity for the time equal to $t=5 \times 10^4 t_g$. The grey dotted line, which corresponds to the same constant value of $l$ as was used to obtain the white line in Fig.~\ref{fig:q=1.07_rcusp=2.3_steady_state_rho}, can be barely distinguished from the solid blue line (to the right of its maximum) representing the data from simulation.

\begin{figure}[!]
\centering 
\includegraphics[width=\columnwidth]{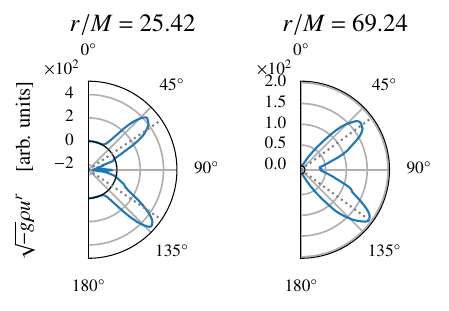}
\caption{Angular distribution of fluxes of rest mass $\sqrt{-g} \mu u^r$ through spherical surfaces of radii $r/M = 25.42$ (outer edge of the initial torus) and $r/M = 69.24$ (a few computational cells from outer boundary) for the $Q=1.07$, $r_\text{cusp}/M=2.30$ simulation. The data are averages over time from $t=10^4 t_g$ to $t=5 \times 10^4 t_g$.}
\label{fig:q=1.02_rcusp=2.3_fluxes}
\end{figure}

Panels in Fig.~\ref{fig:q=1.07_rcusp=2.3_steady_state_equilibrium_curves} present distributions of $l=-u_\phi/u_t$ and the radial velocity component $\varv_r=-u^r/u_t$ at $t=5 \times 10^4 t_g$. The upper panel of Fig.~\ref{fig:q=1.07_rcusp=2.3_steady_state_equilibrium_curves} shows that outflow of material from the initial torus is limited to moderate zenithal angles, and is well separated from the polar axis of the simulation setup at $\theta = 0, \pi$. This observation is also supported by the angular distribution of fluxes presented in Fig.~\ref{fig:q=1.02_rcusp=2.3_fluxes}. At the distance of the maximal radial extension of the initial torus $r/M = 25.42$, the fluxes vanish at high latitudes, with the maximal flux value obtained close to the opening angle of the imaginary shadow of the initial torus cast by the cusp.

The data from a snapshot of our simulation shown in Fig.~\ref{fig:q=1.07_rcusp=2.3_steady_state_equilibrium_curves}, indicate that at the given time the described outflows are not symmetric with respect to the equatorial plane (which is a plane of symmetry in the initial configuration). Only the time average presented in Fig.~\ref{fig:q=1.07_rcusp=2.3_steady_state_rho} displays this symmetry. One may deduce that the outflows alternate between two asymmetric states, analogously to the changes in orientation of the accretion stream, as noted in simulation S02c26 for $Q/M=1.02$.

\begin{figure}[!]
\centering 
\includegraphics[width=\columnwidth]{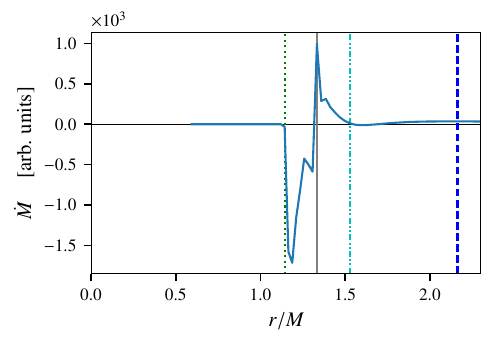}
\caption{The rest mass accretion rate, $\dot{M}$, as a function of the coordinate distance $r/M$ in steady-state accretion, averaged over time from $t=10^4 t_g$ to $t=5 \times 10^4 t_g$. Reissner-Nordström $Q/M=1.07$ naked singularity simulation with the cusp of the initial torus at $r_\text{cusp}/M=2.30$. The vertical lines have the same meaning as in Fig.~\ref{fig:q=1.02_rcusp=2.65_steady_state_accretion_rate}.}
\label{fig:q=1.07_rcusp=2.3_steady_state_accretion_rate}
\end{figure}

Fig.~\ref{fig:q=1.07_rcusp=2.3_steady_state_accretion_rate} presents the accretion rate of the rest mass, $\dot{M}$, as a~function of the coordinate distance $r$ from the naked singularity. The function has root at the location of the center of the inner torus, i.e. maximum of the pressure, depicted by the solid vertical grey line. What distinguishes the results in simulations with $Q/M=1.07$ from the previously considered $Q/M=1.02$ is the fact that the location of the center of the inner torus is significantly farther from the naked singularity than the zero-gravity radius, as can be easily seen in the plots. We can conclude that the accretion rates change sign at the location of the pressure maximum in the inner torus, and not at the location of the zero-gravity sphere, so the accumulated, rotating matter persists in the inner torus for long periods of time.

The second setup studied for $Q/M=1.07$ (with \mbox{$r_\text{cusp}/M=2.50$}) is by our intention similar to the smaller torus considered for $Q/M=1.02$, i.e. the one with location of the cusp at $r_\text{cusp}/M=3.00$.

\begin{figure}[!]
\centering 
\includegraphics[width=\columnwidth]{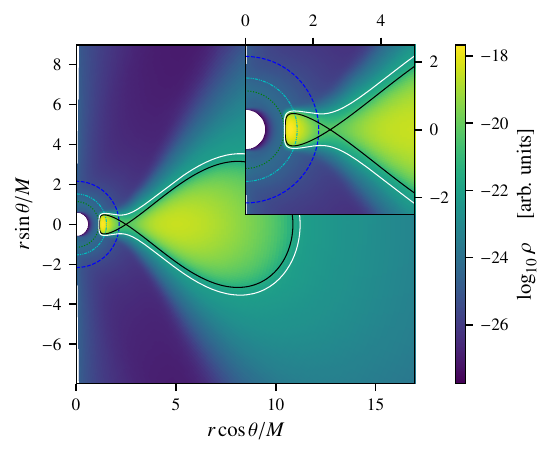}
\caption{Logarithm of rest mass density, $\log_{10} \rho$, in steady-state accretion onto the Reissner-Nordström, $Q/M=1.07$ naked singularity, averaged over time from $t=10^4 t_g$ to $t=5 \times 10^4 t_g$. The meaning of the lines is the same as in Fig.~\ref{fig:q=1.07_rcusp=2.3_steady_state_rho}.}
\label{fig:q=1.07_rcusp=2.5_steady_state_rho}
\end{figure}

\begin{figure}[!]
\centering 
\includegraphics[width=\columnwidth]{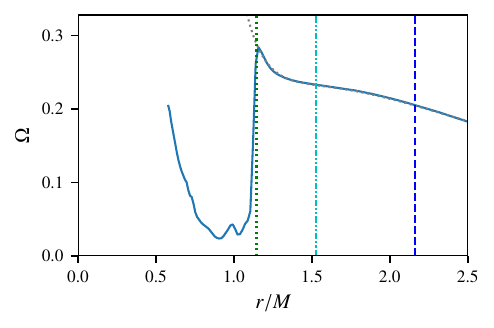}
\caption{Orbital angular frequency, $\Omega(r)$, in the steady state of accretion onto the Reissner-Nordström, $Q/M=1.07$ naked singularity at time $t=5 \times 10^4 t_g$. The cusp of the initial torus was located at $r_\text{cusp}/M=2.50$. The dotted grey line on the main plot is the frequency $\Omega$ for a constant value of the $l$ parameter estimated from the conditions in the inner torus. The green dotted line indicates the zero-gravity sphere radius, the cyan double dotted dashed line the radius of maximum Keplerian frequency, and the dashed blue line the radius of marginally bound orbits.}
\label{fig:q=1.07_rcusp=2.5_steady_state_omega}
\end{figure}

The time averaged rest mass density (from $t= 10^4 t_g$ to $t=5 \times 10^4 t_g$) for the initial torus with cusp located at $r_\text{cusp}/M=2.50$ is presented in Fig.~\ref{fig:q=1.07_rcusp=2.5_steady_state_rho}. The white line depicts the equipotential curve, constructed as discussed previously, from the averaged value of the $l$ parameter inside the inner torus. It turns out that this curve accurately enclose the high density region of the time averaged data. Another hint for the constant value of $l$ parameter inside the inner torus can be found in plot of the angular frequency $\Omega$ presented in Fig.~\ref{fig:q=1.07_rcusp=2.5_steady_state_omega}. The grey dotted line corresponding to the $l=\mathrm{const} $ condition overlaps nearly perfectly with the data from numerical simulations.

\begin{figure}[!]
\centering 
\includegraphics[width=\columnwidth]{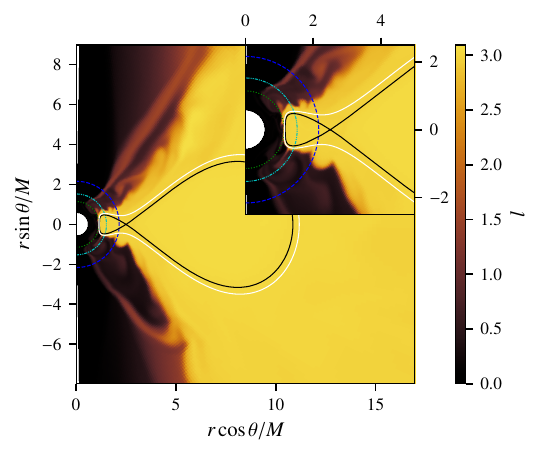}\\
\includegraphics[width=\columnwidth]{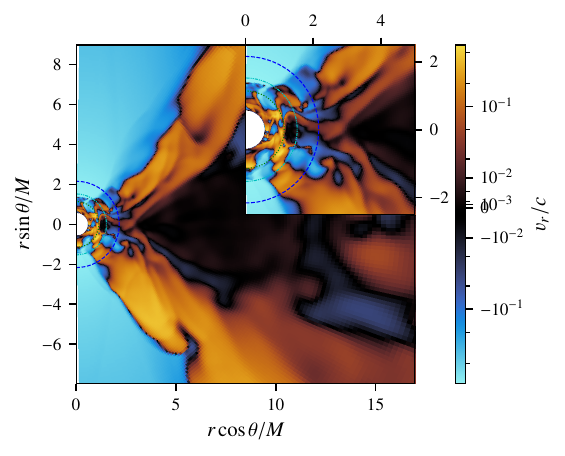}
\caption{A snapshot ($t=5 \times 10^4 t_g$) of steady-state accretion onto the Reissner-Nordström, $Q/M=1.07$ naked singularity. The cusp of the initial torus was located at $r_\text{cusp}/M=2.50$. {\sl Upper panel:} value of $l = -u_\phi / u_t$. 
{\sl Lower panel:} radial component of the velocity, $\varv_r=-u^r/u_t$.
The meaning of the lines is the same as in Fig.~\ref{fig:q=1.07_rcusp=2.3_steady_state_rho}.}
\label{fig:q=1.07_rcusp=2.5_steady_state_equalibrium_curves}
\end{figure}

Panels of Fig.~\ref{fig:q=1.07_rcusp=2.5_steady_state_equalibrium_curves} show the value of the $l = u_\phi / u_t$ parameter and the radial component of the velocity $\varv_r=-u^r/u_t$ from the snapshot from our simulations at time $t=5 \times 10^4 t_g$. The lower panel indicates strong outflows that flow on the surface of the initial torus. The source of the outflowing material can be associated with the initial torus, judging from the upper panel of Fig. \ref{fig:q=1.07_rcusp=2.5_steady_state_equalibrium_curves}, since only the fluid in the initial torus has non-zero azimuthal component of the four-velocity $u^\phi$ and can contribute to non-zero value of the $l$ parameter.

In contrast to the previously considered cases, for \mbox{$Q/M=1.07$} with $r_\text{cusp}/M=2.50$, the symmetry of reflections with respect to the equatorial plane is not fully restored in time averaged data from our simulation. The angular distributions of fluxes $\sqrt{-g} \rho u^r$, $\sqrt{-g}T^r_t$, $\sqrt{-g}T^r_\phi$ are asymmetric with respect to the equatorial plane ($\theta=\pi/2$) even though they are averages over time from $t=10^4t_g$ up to $t=5 \times 10^4t_g$. We expect that the time of alternation between opposite orientations of outflows for this simulation is longer than for previous ones and the $4 \times 10^4t_g$ interval over which the averages are computed is too short to include many transitions. It is worth pointing that the maxima of fluxes at the coordinate distance of $r/M=10.78$ are close to the opening angle of the imaginary shadow of the initial torus cast by the cusp.

\begin{figure}[!]
\centering 
\includegraphics[width=\columnwidth]{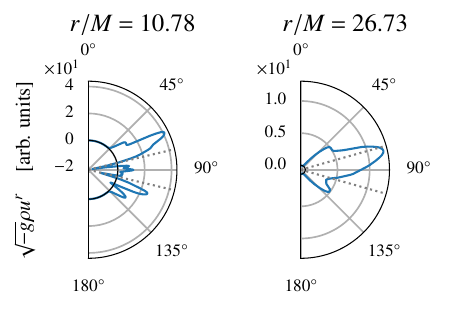}
\caption{Angular distribution of rest-mass flux, $\sqrt{-g} \rho u^r$ through spherical surfaces with radii $r/M = 10.78$ (outer edge of the initial torus) and $r/M = 26.73$ (a few computational cells from outer boundary) for the Reissner-Nordström, $Q=1.07$ simulation. The cusp of the initial torus was located at $r_\text{cusp}/M=2.50$. The data are averages over time from $t=10^4 t_g$ to $t=5 \times 10^4 t_g$.}
\label{fig:q=1.07_rcusp=2.5_fluxes}
\end{figure}

\begin{figure}[!]
\centering 
\includegraphics[width=\columnwidth]{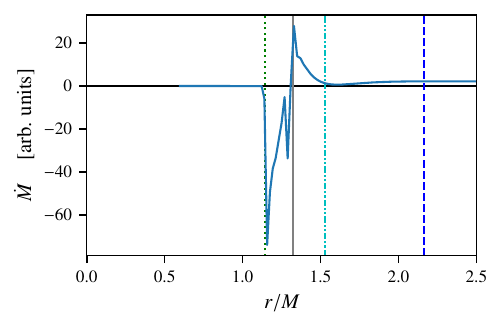}
\caption{Rest-mass accretion rate, $\dot M$, in the steady state of accretion onto the Reissner-Nordström $Q/M=1.07$ naked singularity as a function of the coordinate distance, at time $t=5 \times 10^4 t_g$. The cusp of the initial torus is located at $r_\text{cusp}/M=2.50$. The vertical lines have the same meaning as in Fig.~\ref{fig:q=1.02_rcusp=2.65_steady_state_accretion_rate}.}
\label{fig:q=1.07_rcusp=2.5_steady_state_accretion_rate}
\end{figure} 

\subsection{Case study: $Q/M=1.09$}
The purpose of considering the case $Q/M=1.09$ is to investigate the influence of the (non)existence of unbound circular orbits on the process of accretion onto naked singularities. For $Q/M=1.09$, all circular orbits down to the zero-gravity sphere in the spacetime of the Reissner-Nordström metric have energies smaller than the test particle at an infinite distance, so for this value of the charge to mass ratio marginally bound circular orbits do not exist.

We performed two simulations for $Q/M=1.09$, initialized with equilibrium tori with $r_\text{cusp}/M = 2.00$ (S09c20) and $r_\text{cusp}/M = 2.30$ (S09c23) that were targeted to be analogues of situations discussed in Section~\ref{sec:Q=1.02} for $Q/M=1.02$, with $r_\text{cusp}/M = 2.50$ (S02c26) and $r_\text{cusp}/M = 3.00$ (S02c30) respectively. The drawback of this choice of locations of cusps in the $Q/M=1.09$ simulations is that the inner lobes of the self-intersecting equipotential surfaces are very small, consequently they are quickly filled by the accreted matter in the course of the simulation.

\begin{figure}[!]
\centering 
\includegraphics[width=\columnwidth]{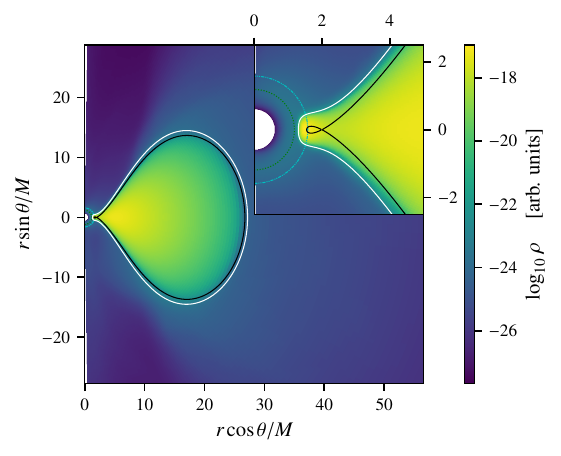}
\caption{Steady state of accretion onto the naked singularity of the Reissner-Nordström spacetime with $Q/M=1.09$. The plot presents the logarithm of rest mass density $\log_{10} \rho$ averaged over time from $t=10^4 t_g$ to $t=5 \times 10^4 t_g$. Black lines correspond to the equipotential surface of the initial torus. White line corresponds to surface of constant potential, with value equal to that of the initial torus surface, with the parameter $l_0$ computed as an average of $l$ over the highly dense region close to naked singularity. The green dotted line correspond to zero-gravity sphere and the cyan double-dotted dashed line is the location of maximum of Keplerian frequency $\Omega_\mathrm{K}$.}
\label{fig:q=1.09_rcusp=2._steady_state_rho}
\end{figure}

\begin{figure}[!]
\centering 
\includegraphics[width=\columnwidth]{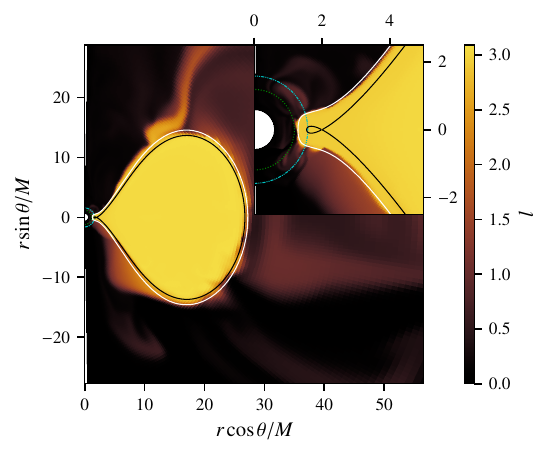}\\
\includegraphics[width=\columnwidth]{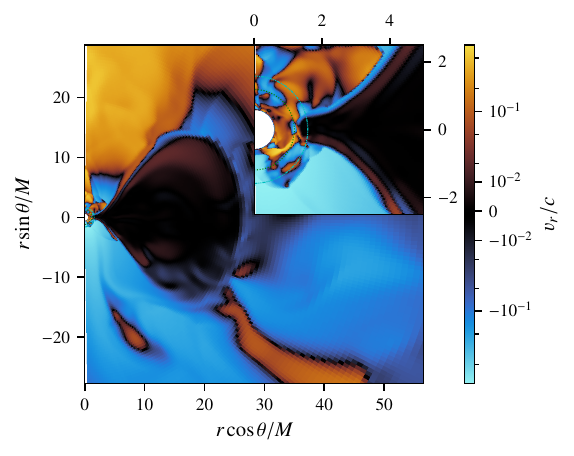}
\caption{Steady-state of accretion with $r_\text{cusp}/M=2.00$ onto the Reissner-Nordström, $Q/M=1.09$ naked singularity. The data are a snapshot of the simulation results at time $t=5 \times 10^4 t_g$. {\sl Upper panel:} value of $l =- u_\phi / u_t$.
{\sl Lower panel:} radial component $\varv_r=-u^r/u_t$ of velocity. 
The zero-gravity sphere is shown with the green dotted line, Keplerian frequency attains a maximum at the radius of the cyan double dotted dashed line.}
\label{fig:q=1.09_rcusp=2._steady_state_equilibrium_curves}
\end{figure}

The  mass energy density $\rho$ averaged over time from \mbox{$t=10^4\ t_g$} to $t=5 \times 10^4\ t_g$ in the simulation with $Q/M=1.09$ and the initial torus with the cusp located at $r_\text{cusp}/M = 2.00$   is presented in Fig.~\ref{fig:q=1.09_rcusp=2._steady_state_rho}; the $l$ parameter and the radial component $\varv_r$ of the velocity of the flow at $t=5 \times 10^4\ t_g$ can be inspected in Fig.~\ref{fig:q=1.09_rcusp=2._steady_state_equilibrium_curves}. In this simulation (S09c20) the inner torus unites with the outer one forming a single structure. 

\begin{figure}[!]
\centering 
\includegraphics[width=\columnwidth]{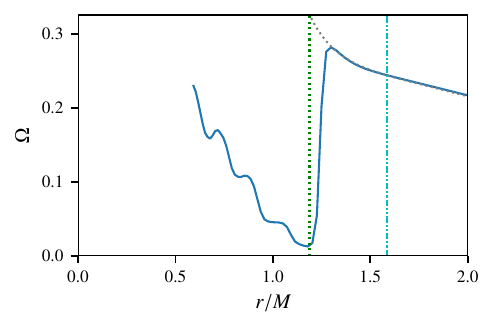}
\caption{Orbital angular frequency as a function of the coordinate distance, $\Omega(r)$, at time $t=5 \times 10^4 t_g$ in the steady state of accretion onto the Reissner-Nordström $Q/M=1.09$ naked singularity. The cusp of the initial torus was located at $r_\text{cusp}/M=2.00$. The dotted grey line on the main plot is the frequency $\Omega$ for a constant value of the $l$ parameter estimated from the conditions in the inner torus. The green dotted line indicates the zero-gravity sphere radius, the cyan double dotted dashed line the radius of maximum Keplerian frequency.}
\label{fig:q=1.09_rcusp=2._steady_state_omega}
\end{figure}

Even though the mass-density based masking of the inner torus is ambiguous in this case, our procedure gives an equipotential curve which encloses the mass density distribution of the inner torus quite precisely. The orbital frequency profile $\Omega(r)$ calculated from the $l$ parameter value inside the inner torus reproduce the data from simulations precisely, as is presented in Fig.~\ref{fig:q=1.09_rcusp=2._steady_state_omega}.

\begin{figure}[!]
\centering 
\includegraphics[width=\columnwidth]{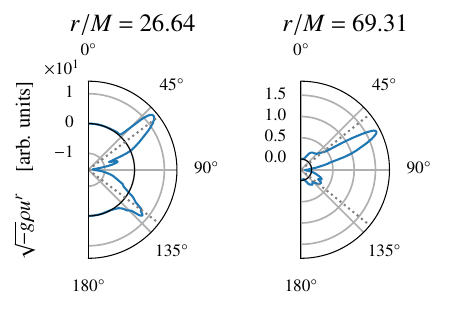}
\caption{Angular distribution of rest-mass flux $\sqrt{-g} \rho u^r$ through spherical surfaces with radii $r/M = 26.64$ (outer edge of the initial torus) and $r/M = 69.31$ (a few computational cells from outer boundary) in a simulation for the Reissner-Nordström, $Q=1.09$ naked singularity. The cusp of the initial torus was located at $r_\text{cusp}/M=2.00$. The data are averages over time from $t=10^4 t_g$ to $t=5 \times 10^4 t_g$.}
\label{fig:q=1.09_rcusp=2._fluxes}
\end{figure}

As in the previously described simulations, the high mass density inflow of matter from the outer torus takes place along the surface of the torus and is accompanied by a lower density outflow outside the torus. The outflow of the rest-mass density $\rho$ (also energy density and angular momentum density) is collimated above the surface of the outer torus. The angular distribution of the flux $\sqrt{-g} \rho u^r$ of the rest-mass density, presented in Fig.~\ref{fig:q=1.09_rcusp=2._fluxes}, shows asymmetry of outflows with respect to the equatorial plane.

\begin{figure}[!]
\centering 
\includegraphics[width=\columnwidth]{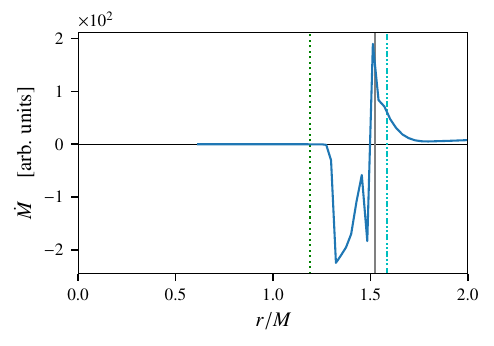}
\caption{Rest-mass accretion rate, $\dot M$, in the steady state of accretion onto the Reissner-Nordström $Q/M=1.09$ naked singularity as a function of the coordinate distance, at time $t=5 \times 10^4 t_g$. The cusp of the initial torus was located at $r_\text{cusp}/M=2.30$. The vertical lines have the same meaning as in Fig.~\ref{fig:q=1.02_rcusp=2.65_steady_state_accretion_rate}.}
\label{fig:q=1.09_rcusp=2._steady_state_accretion_rate}
\end{figure}

Figs.~\ref{fig:q=1.09_rcusp=2._steady_state_rho}, and \ref{fig:q=1.09_rcusp=2._steady_state_equilibrium_curves}
show that matter accumulates inside the inner torus with its circle of maximum pressure well outside the zero-gravity radius. A plot of rest mass accretion rate, presented in Fig.~\ref{fig:q=1.09_rcusp=2._steady_state_accretion_rate}, proves that the accreting matter continues to accumulate there at late times of the simulation. For the simulated setups with $Q/M=1.09$ we can confirm that most of the matter orbits the naked singularity at some distance from the zero-gravity sphere.

\begin{figure}[!]
\centering 
\includegraphics[width=\columnwidth]{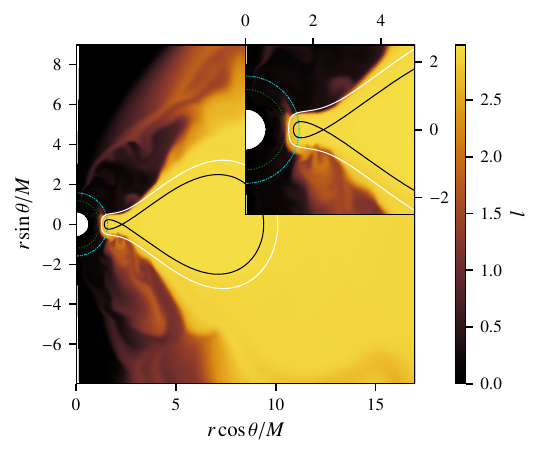}\\
\includegraphics[width=\columnwidth]{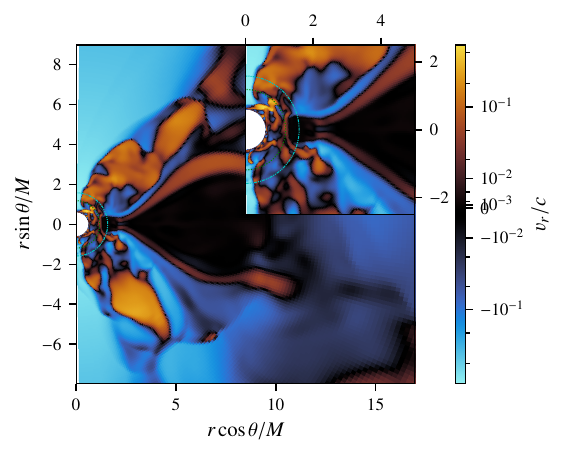}
\caption{A snapshot at $t=5 \times 10^4 t_g$ of steady-state accretion onto the Reissner-Nordström, $Q/M=1.09$ naked singularity. 
{\sl Upper panel:} value of $l = -u_\phi / u_t$. 
{\sl Lower panel:} radial component of the velocity, $\varv_r=-u^r/u_t$. Black lines correspond to the equipotential surface of the initial torus. White lines correspond to equipotential surfaces with value of $W$ equal to that of the initial torus, and the parameter $l_0$ computed as an average of $l$ over the highly dense region close to naked singularity. The green dotted line corresponds to zero-gravity sphere, the cyan double dotted dashed line to the radial location of maximum of Keplerian frequency.}
\label{fig:q=1.09_rcusp=2.3_steady_state_equilibrium_curves}
\end{figure}

\begin{figure}[!]
\centering 
\includegraphics[width=\columnwidth]{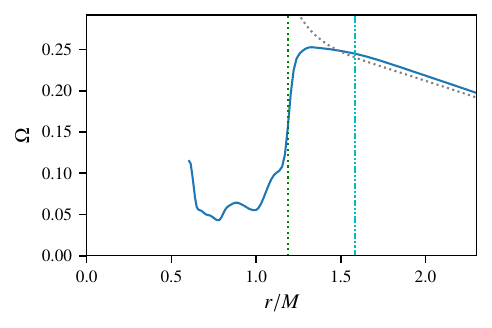}
\caption{Orbital angular frequency, $\Omega(r)$, in the steady state of accretion onto the Reissner-Nordström $Q/M=1.09$ naked singularity at time $t=5 \times 10^4 t_g$. The cusp of the initial torus was located at $r_\text{cusp}/M=2.30$. The dotted grey line is the frequency $\Omega$ for a constant value of $l$ estimated from the conditions in the inner torus. The green dotted line indicates the zero-gravity sphere radius, the cyan double dotted dashed line the radius of maximum Keplerian frequency.}
\label{fig:q=1.09_rcusp=2.3_steady_state_omega}
\end{figure}

More severe problems with masking the inner torus occur in the setup of the second simulation (S09c23) for \mbox{$Q/M=1.09$}. The inner and outer tori merge quite fast in this case, and they cannot be unambiguously distinguished, one from the other. As is visible in Fig.~\ref{fig:q=1.09_rcusp=2._steady_state_rho}, the proposed equipotential curve fits the high mass density region moderately well. Even though it encloses nearly perfectly the high $l$ region close to the naked singularity presented in the upper panel of Fig.~\ref{fig:q=1.09_rcusp=2.3_steady_state_equilibrium_curves}, the hypothesis of constant $l$ does not reproduce the angular frequency $\Omega(r)$ of the fluid as well as in the previous cases (Fig.~\ref{fig:q=1.09_rcusp=2.3_steady_state_omega}).

\begin{figure}[!]
\centering 
\includegraphics[width=\columnwidth]{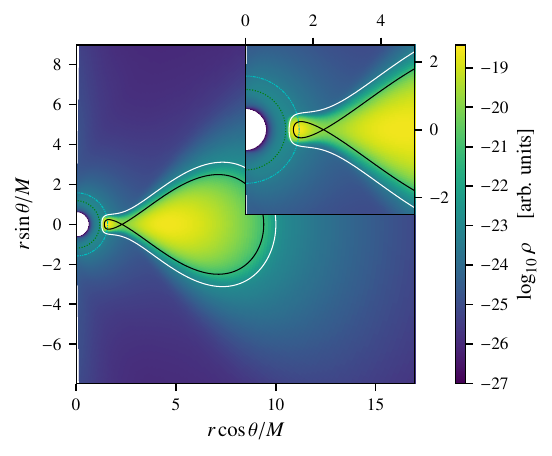}
\caption{Logarithm of rest mass density $\log_{10} \rho$, averaged over time from $t=10^4 t_g$ up to $t=5 \times 10^4 t_g$, in steady state of the accretion onto the Reissner-Nordström, $Q/M=1.09$ naked singularity. Black lines correspond to the equipotential surface of the initial torus. White lines correspond to equipotential surfaces with value of $W$ equal to that of the initial torus, and the parameter $l_0$ computed as an average of $l$ over the highly dense region close to naked singularity. The green dotted line indicates the zero-gravity sphere, the cyan double dotted dashed line the maximum of Keplerian frequency.}
\label{fig:q=1.09_rcusp=2.3_steady_state_rho}
\end{figure}

\begin{figure}[!]
\centering 
\includegraphics[width=\columnwidth]{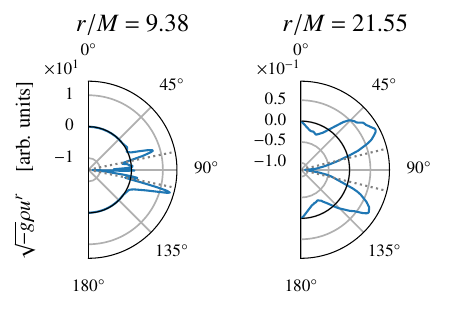}
\caption{Angular distribution of rest-mass flux, $\sqrt{-g} \rho u^r$, through spherical surfaces with radii $r/M = 9.38$ (outer edge of the initial torus) and $r/M = 21.55$ (a few computational cells from outer boundary) for a Reissner-Nordström, $Q=1.09$ simulation with the initial torus cusp located at $r_\text{cusp}/M=2.30$. The data are averaged over time from $t=10^4 t_g$ to $t=5 \times 10^4 t_g$.}
\label{fig:q=1.09_rcusp=2.3_fluxes}
\end{figure}

As for the torus with $r_\text{cusp}/M = 3.00$ in $Q/M=1.02$ and $r_\text{cusp}/M = 2.50$ in $Q/M=1.07$ cases, an outflow of moderate density material is observed outside the outer torus at fairly large polar angles, as visible in Fig.~\ref{fig:q=1.09_rcusp=2.3_fluxes}. Interestingly, for the case of $Q/M=1.09$ with $r_\text{cusp}/M = 2.30$, the outflows averaged in time from $t=10^4 t_g$ to $t=5 \times 10^4 t_g$ are nearly symmetric with respect to the equatorial plane.

\begin{figure}[!]
\centering 
\includegraphics[width=\columnwidth]{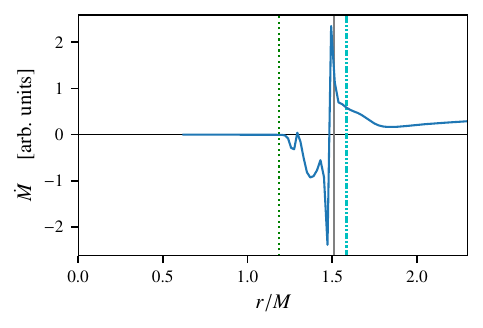}
\caption{Rest-mass accretion rate, $\dot M$, as a function of the coordinate distance in the steady state of accretion onto the Reissner-Nordström $Q/M=1.09$ naked singularity, averaged over time from $t=10^4 t_g$ up to $t=5 \times 10^4 t_g$. 
The cusp of the initial torus was located at $r_\text{cusp}/M=2.30$.
The vertical lines have the same meaning as in Fig.~\ref{fig:q=1.02_rcusp=2.65_steady_state_accretion_rate}.}
\label{fig:q=1.09_rcusp=2.3_steady_state_accretion_rate}
\end{figure}

In the case of $Q=1.09$ with $r_\text{cusp}/M = 2.30$, as for the others discussed simulations, the accreting matter keeps accumulating around the center of the inner torus, as is visible in the plots of angular momentum accretion rate, Fig.~\ref{fig:q=1.09_rcusp=2.3_steady_state_accretion_rate}, where this location is denoted with a vertical grey line.

\section{Conclusions\label{sec:summary}}
We investigated accretion of electrically neutral fluid onto naked singularities described by the Reissner-Nordström metric with charge $Q$ greater than the mass $M$. In our studies, we used the GR hydrodynamical numerical code \texttt{Koral} \citep{Sadowski:2012} to solve for dynamics of a perfect fluid accreting onto naked singularities of astrophysically interesting macroscopic size. 

Observations of the core regions of the nucleus of our own Galaxy and of M87, made possible by the gravitational radius scale resolution of the Event Horizon Telescope \citep{EHT5,EHT2022}, gave a new impetus for the assessment of observational differences between the supermassive black holes and naked singularities. Any additional structures developing in the vicinity of the naked singularity would, naturally, introduce new features which could leave an imprint in the observational record.

In our earlier paper \citep{Kluzniak:2024cxm}, we investigated the difference between the accretion onto black holes and naked singularities described by the Reissner-Nordström metric. We have found that the fundamental difference in the existence, or not, of the event horizon has profound consequences for observational possibilities of distinguishing astrophysical accreting naked singularities from black holes. Our simulations revealed that the flow is completely different in the two cases, with the matter being absorbed by black holes, but ejected in powerful outflows by naked singularities. Unlike the well-studied scenario of black holes in which the accreting matter falls under the event horizon and cannot subsequently be observed by a~distant observer, the matter falling onto naked singularities cannot be hidden inside horizon and may be observed even after having accumulated in the form of a rotating toroidal body in the vicinity of the zero-gravity sphere. 
Another difference is that some of the fluid falling towards the naked singularity is reflected and may form powerful outflows. In \cite{Kluzniak:2024cxm}, we have speculated that our findings may be particularly relevant to the ultrafast outflows (UFOs) from AGNs, which are difficult to explain in conventional models \citep{Keshet:2022gxs}. 

In the current paper we focused on accretion onto RN naked singularities. To better understand our numerical findings we discussed our simulation results in the context of analytically derived equipotential surfaces, for suitable values of the specific angular momentum parameter $l$. We also provided further details of the properties of outflows produced during accretion onto naked singularities described by the Reissner-Nordström metric in the regime of \mbox{$1 < Q/M < \sqrt{5} / 2$}, including its parametric dependence on $Q/M$.

In our naked singularity simulation, a toroidal structure is formed from the accreted material near the zero-gravity sphere. We have found that the quantity $l = u_\phi / u_t$, which we initially set to a constant, $l=l_0$, as the parameter defining the initial torus, during accretion takes a smaller, but still approximately uniform value inside the inner toroidal structure. We ascribe this change of the value of $l$ to the heating of the material during accretion. 
In the absence of viscous torques, the angular momentum of the fluid is conserved during the flow. 
However, the proper angular momentum density, $j = w u^t u_\phi$, depends not only on the components of the four-velocity of the fluid $u$, but also on the temperature, through the enthalpy $w$. When the latter increases with heating of the fluid the azimuthal component of the four-velocity $u_\phi$ needs to decrease to keep the angular momentum constant. 

The heated matter might pose a~significant source of radiation emitted by the system of the naked singularity, not included in previous analytic studies. Furthermore, the accumulated matter might form an optically thick structure which will significantly influence the propagation of the radiation in the vicinity of the naked singularity. In the recent studies of \citet{Tavlayan:2023vbv,daSilva:2023jxa,Mummery:2024znv,Viththani:2024fod} this effect was not recognized. The conclusion that an~image of flow towards the naked singularity should be characterized by the additional contribution which in the case of black holes is not visible, because it falls under the horizon may have to be revisited, as 
the toroidal structure of accreted matter might obscure a certain fraction of the emitted radiation. A detailed study of this problem would require the ray-tracing computations which are beyond the scope of this paper and are postponed to the future research.

Some fraction of the material falling onto the naked singularity will be ejected in a strong outflow. This flow of the plasma from naked singularity is not collimated in poloidal regions as jets observed in magneto-hydrodynamical simulations of accretion onto black holes, but is concentrated at intermediate polar angles. We have found that the geometry of outflows is dictated by the shape of the initial torus used as a~source of the material in our simulations. The outflowing matter is pushed onto the surface of the outer torus and slides over its surface (in the opposite direction to the matter inside the torus which accretes toward the naked singularity under the surface of the torus), propagating beyond the extension of the torus, eventually diffusing and bending toward the equatorial plane.

It would be instructive to incorporate the dynamics of the magnetic field, since it is known, from studies of accretion onto black holes, that the magnetic field of proper topology can collimate outflows into jets. In magnetohydrodynamical simulations the observed pattern of angular distribution of outflows can be modified, making them more similar to observed astrophysical jets. However, performing such simulations requires sustaining the consistency of the numerical scheme with arbitrary high magnetization in vicinity of the naked singularity, where we expect a vacuum to form. We postpone such studies till our code is developed to a point at which it can reliably simulate the systems in question.

\begin{acknowledgements}
Research supported in part by the Polish National Science Centre grants No.~2019/35/O/ST9/03965 and 2019/33/B/ST9/0156. During completion of this manuscript T.K. was supported by Polish National Science Centre grant no 2023/51/B/ST9/00943. High-performance computations was performed on CHUCK cluster in CAMK Warsaw. The authors thank Debora Lan\v{c}ov\'a, Daniela Pugliese and Miljenko \v{C}emelji\'{c} for stimulating discussions. We thank Ruchi Mishra and Angelos Karakonstantakis for testing the numerical code in early stages of this work. 
\end{acknowledgements}

\bibliographystyle{aa}
\bibliography{references}

\begin{appendix}

\section{Numerical methods used in \texttt{Koral+} code\label{sec:numerical_code}}
Results presented in Section \ref{sec:numerical_results} were obtained using code called \texttt{Koral+} which is an~extension of well-known \texttt{Koral} \cite{Sadowski:2012, Sadowski:2013gua}. The used version of the code is the intermediate step in ongoing, long-term project of modernizing the \texttt{Koral} legacy code. Up to now the the metric dependent part of the code was refactorized and the new part simplifying implementation of new systems of coordinates exploiting symbolic computations software was added.

Presented results of numerical simulations were computed in modified Boyer-Linquist coordinates which has radial coordinate $s$ logarithmically stretched, i.e. $s = \log{r}$ with respect to standard Boyer-Linquist coordinates $(t,r,\theta,\phi)$.

The rest of used numerical methods follow from \texttt{Koral} code and have been already described in details in \cite{Sadowski:2012, Sadowski:2013gua}. We will only briefly present our numerical setup.

Equations of relativistic hydrodynamics can be derived from conservation 
\begin{equation}
    \nabla_\mu T^{\mu \nu} = 0 \label{eq:energy_momentum_conservation}
\end{equation}
of energy-momentum tensor of the perfect fluid
\begin{equation}
    T^{\mu \nu} = w u^\mu u ^\nu + p g^{\mu \nu}
\end{equation}
which is usually in astrophysical context supplemented with rest mass conservation
\begin{equation}
    \nabla_\mu \left( \rho u^\mu \right) = 0. \label{eq:rest_mass_conservation}
\end{equation}
\texttt{Koral(+)} solves Eqs. \eqref{eq:energy_momentum_conservation} and \eqref{eq:rest_mass_conservation} in the form
\begin{subequations}
    \label{eq:conservation_form}
    \begin{align}
        \partial_t T^{t \nu} + \partial_i T^{i \nu} = -\Gamma^{\nu}_{\phantom{\nu} \mu \lambda} T^{\mu \lambda} - \frac{1}{\sqrt{-g}} \partial_i \left(\sqrt{-g}\right) T^{i \nu},\\
        \partial_t \left( \rho u^t\right) + \partial_i \left( \rho u^i \right) = - \frac{1}{\sqrt{-g}} \partial_i \left(\sqrt{-g}\right) \left( \rho u^i \right)
    \end{align}
\end{subequations}
which is accessible to standard Godunov-type numerical schemes for conservation equations.

\texttt{Koral(+)} is based on method of lines in which the equations \eqref{eq:conservation_form} are first discretized in space and then integrated in time using time stepping approach. The spacial discretization starts with conversion of so called conserved quantities:\footnote{In this manuscript we describe only hydrodynamical part of the code. However, \texttt{Koral+} as its ancestor \texttt{Koral} is able to solve also equations of magneto-hydrodynamics coupled to radiation using M1 closer scheme for the later.}
\begin{equation}
    U = [\rho u^t, T^t_{\phantom{t}t} + \rho u^t, T^t_{\phantom{t}i}, S u^t]
\end{equation}
to so called primitive ones:
\begin{equation}
    P = [\rho, \varepsilon, u^i, S],
\end{equation}
where $S$ is the entropy of the fluid which satisfies the following conservation equation
\begin{equation}
    \nabla_\mu \left( S u^\mu \right) = 0. \label{eq:entropy_conservation}
\end{equation}
Set of equations \eqref{eq:energy_momentum_conservation} and \eqref{eq:rest_mass_conservation} supplemented with \eqref{eq:entropy_conservation} is over-specified and \texttt{Koral(+)} solve \eqref{eq:entropy_conservation} only to use evolved entropy in backup procedure \cite{Sadowski:2013gua} of obtaining primitive variables from conserved ones if the default algorithm "$1D_W$" of \cite{Noble:2005gf} fails. In opposite case when conversion based on $\rho u^t$, $T^t_{\phantom{t}t}$, $T^t_{\phantom{t}i}$ succeed the entropy is calculated from $\rho$ and $p$ as
\begin{equation}
    S = \frac{\rho}{\Gamma - 1} \log{\left( \frac{p(\varepsilon)}{\rho^\Gamma} \right)}
\end{equation}
and evolved up to the end of timestep.

After the conversion the obtained primitive variables are corrected up to assumed numerical floors. The most important for our simulations are minimal value of $\rho$ which we assume is $10^{-50}$ is simulation units. This value is usually set in the interior of zero-gravity sphere as described in Section~\ref{sec:numerical_results}. Furthermore, the diluted plasma contained inside zero-gravity sphere is easily heated up by the accreting material which is limited by the maximal $\varepsilon / \rho$ ratio equal to $100$. Motivated by the findings of \cite{Siegel:2017sav} we limited maximal relativistic $\gamma = u^t / \sqrt{-g^{tt}}$ to $10$. This choice influence the dynamics of the relaxation of initial conditions inside the zero-gravity sphere, but is not important for later process of accretion of material from torus. Certain finite volumes during performed simulations are prone to obtain negative internal density $\varepsilon$ during conversion step. We decided to fix such pathological cells by substituting problematic values by the averages from neighbouring cells of the computational mesh. Due to high resolution (especially around zero-gravity sphere where such situation takes place) this fix-up procedure does not spoil the physical results obtained in the simulations.

The conversion from conserved quantities to primitive ones is followed by the reconstruction of quantities located at faces of finite volumes from volume averages. This means that \texttt{Koral(+)} reconstructs the primitive quantities and not the conserved ones. In our simulations we used linear interpolation for reconstruction corrected by the van-Leer’s minmod limiter with the dissipation parameter $\theta_\text{minmod} = 1.5$ \citep{Kurganov:2000241}. After limiting, the numerical floors are applied to reconstructed quantities.

Reconstructed primitive variables are used to calculate fluxes using approximate Riemann solvers. For this research project we chose Harten, Lax and van Leer (HLL) solver \cite{Harten:1983}.

Calculated fluxes and sources from right hand sides of \eqref{eq:conservation_form} are used to integrate conservation equations in time by midpoint method which is of second order in timestep length.

\section{Locations of marginally bound orbits\label{sec:MBO_roots}}

Solutions of equation \eqref{eq:MBO_equation} for $32 M^2 - 27 Q^2 > 0$ take the form:
\begin{align}
    r_\text{MBO}^- &= \frac{4}{3} M + \frac{4}{3} \sqrt{4 M^2 - 3 Q^2}\\
    &\times \cos{\left[\frac{1}{3} \arccos{\left(\frac{128 M^4 - 144 M^2 Q^2 + 27 Q^4}{16 M (4 M^2 - 3 Q^2)^{3/2}} \right)} \right]}, \\
    r_\text{MBO}^+ &= \frac{4}{3} M - \frac{4}{3} \sqrt{4 M^2 - 3 Q^2}\\
    &\times\sin{\left[\frac{1}{3} \arcsin{\left(\frac{128 M^4 - 144 M^2 Q^2 + 27 Q^4}{16 M (4 M^2 - 3 Q^2)^{3/2}} \right)} \right]}, \\
    r_\text{MBO}^c &= \frac{4}{3} M - \frac{4}{3} \sqrt{4 M^2 - 3 Q^2}\\
    &\times\sin{\left[\frac{\pi}{6} + \frac{1}{3} \arccos{\left(\frac{128 M^4 - 144 M^2 Q^2 + 27 Q^4}{16 M (4 M^2 - 3 Q^2)^{3/2}} \right)} \right]},
\end{align}
with $r_\text{MBO}^c$ located below zero-gravity radius $r_0$.

\section{Supplementary results\label{sec:supplemental}}
In this section, we provide further results from simulations which
support our hypothesis, but are qualitatively analogous to cases discussed 
in Section \ref{sec:numerical_results}. We believe that they can be useful for other researchers working in the field.

\subsection{Angular distribution of fluxes}
\begin{minipage}[t]{0.48\columnwidth}
\begin{figure}[H]
\centering 
\includegraphics[width=\columnwidth]{Figs/Q=1.02_RCUSP=3._10000_50000_polar_rho}\\
\includegraphics[width=\columnwidth]{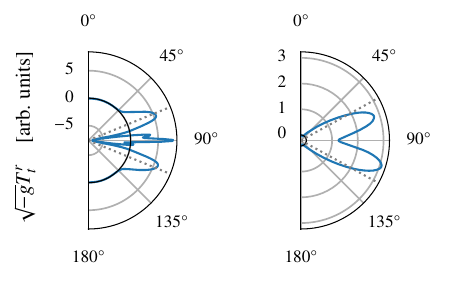}\\
\includegraphics[width=\columnwidth]{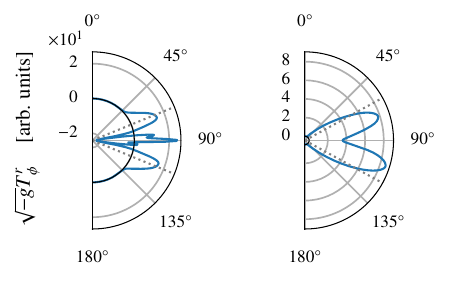}
\caption{Angular distribution of fluxes of rest mass $\sqrt{-g} \rho u^r$ (top panel), energy $\sqrt{-g} T^r_t$ (middle panel) and angular momentum $\sqrt{-g} T^r_\phi$ (bottom panel) through the spherical surfaces with radii $r/M =8.66$ (outer edge of the initial torus) and $r/M = 21.47$ (a~few computational cells from outer boundary) for the Reissner-Nordström, $Q=1.02$ simulation with the cusp of the initial torus located at $r_\text{cusp}/M=3.00$. The data are time averages of the simulation results from $t=10^4 t_g$ to $t=5 \times 10^4 t_g$.}
\end{figure}
\end{minipage}
\hfill
\begin{minipage}[t]{0.48\columnwidth}
\begin{figure}[H]
\centering 
\includegraphics[width=\columnwidth]{Figs/Q=1.07_RCUSP=2.3_10000_50000_polar_rho}\\
\includegraphics[width=\columnwidth]{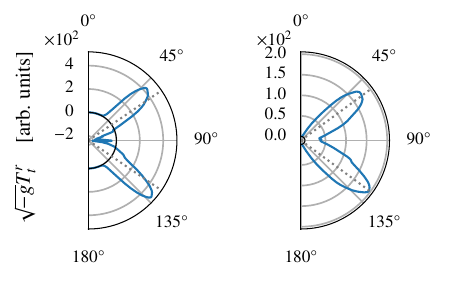}\\
\includegraphics[width=\columnwidth]{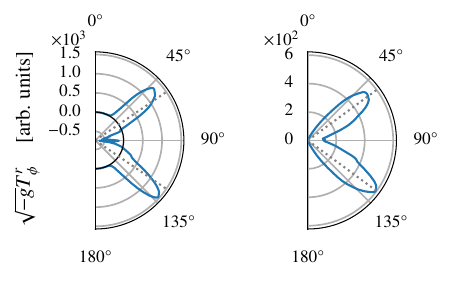}
\caption{Angular distribution of fluxes of rest mass $\sqrt{-g} \rho u^r$ (top panel), energy $\sqrt{-g} T^r_t$ (middle panel) and angular momentum $\sqrt{-g} T^r_\phi$ (bottom panel) through the spherical surfaces with radii $r/M = 25.42$ (outer edge of the initial torus) and $r/M = 69.24$ (a~few computational cells from outer boundary) from simulation for the Reissner-Nordström, $Q=1.07$ simulation with the cusp of the initial torus located at $r_\text{cusp}/M=2.30$. The data are time averages of the simulation results from $t=10^4 t_g$ to $t=5 \times 10^4 t_g$.}
\end{figure}
\end{minipage}

\begin{minipage}[t]{0.48\columnwidth}
\begin{figure}[H]
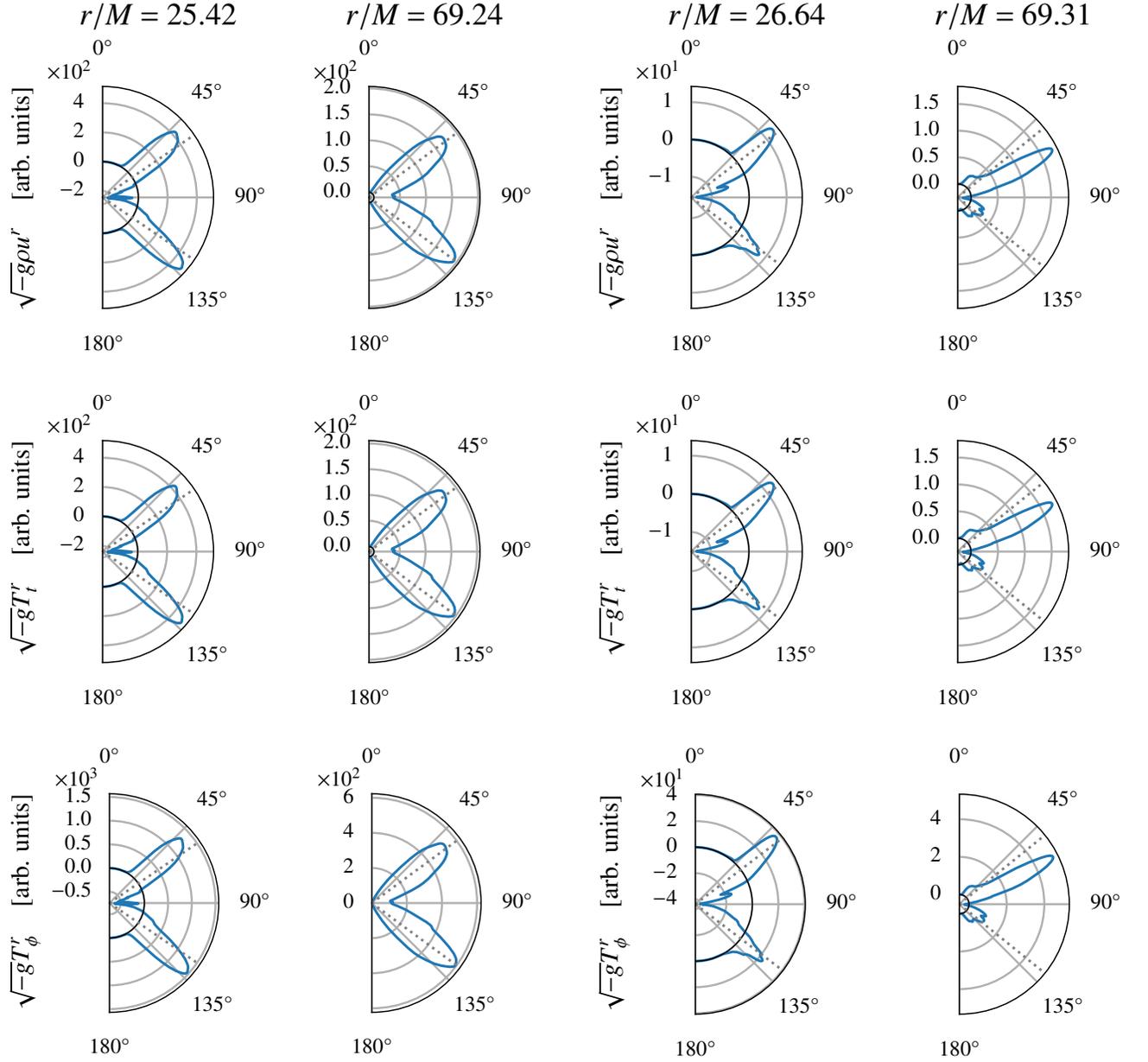

\centering 
\includegraphics[width=\columnwidth]{Figs/Q=1.07_RCUSP=2.3_10000_50000_polar_rho}\\
\includegraphics[width=\columnwidth]{Figs/Q=1.07_RCUSP=2.3_10000_50000_polar_E}\\
\includegraphics[width=\columnwidth]{Figs/Q=1.07_RCUSP=2.3_10000_50000_polar_Lz}
\caption{Angular distribution of fluxes of rest mass $\sqrt{-g} \rho u^r$ (top panel), energy $\sqrt{-g} T^r_t$ (middle panel) and angular momentum $\sqrt{-g} T^r_\phi$ (bottom panel) through the spherical surfaces with radii $r/M = 25.42$ (outer edge of the initial torus) and $r/M = 69.24$ (a~few computational cells from outer boundary) from simulation for the Reissner-Nordström, $Q=1.07$ simulation with the cusp of the initial torus located at $r_\text{cusp}/M=2.30$. The data are time averages of the simulation results from $t=10^4 t_g$ to $t=5 \times 10^4 t_g$.}
\end{figure}
\end{minipage}
\hfill
\begin{minipage}[t]{0.48\columnwidth}
\begin{figure}[H]
\centering 
\includegraphics[width=\columnwidth]{Figs/Q=1.09_RCUSP=2._10000_50000_polar_rho}
\includegraphics[width=\columnwidth]{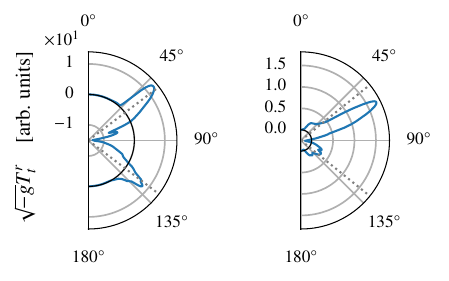}
\includegraphics[width=\columnwidth]{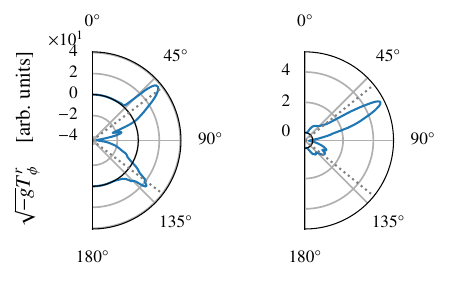}
\caption{Angular distribution of fluxes of rest mass $\sqrt{-g} \rho u^r$ (top panel), energy $\sqrt{-g} T^r_t$ (middle panel) and angular momentum $\sqrt{-g} T^r_\phi$ (bottom panel) through the spherical surfaces with radii $r/M = 26.64$ (outer edge of the initial torus) and $r/M = 69.31$ (a~few computational cells from outer boundary) from simulation for the Reissner-Nordström metric with $Q=1.09$. The cusp of the initial torus is located at $r_\text{cusp}/M=2.00$. The data are time averages of the simulation results from $t=10^4 t_g$ to $t=5 \times 10^4 t_g$.}
\end{figure}
\end{minipage}

\begin{minipage}[t]{0.48\columnwidth}
\begin{figure}[H]
\centering 
\includegraphics[width=\columnwidth]{Figs/Q=1.09_RCUSP=2.3_10000_50000_polar_rho}\\
\includegraphics[width=\columnwidth]{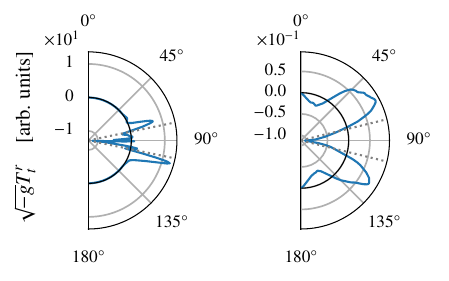}\\
\includegraphics[width=\columnwidth]{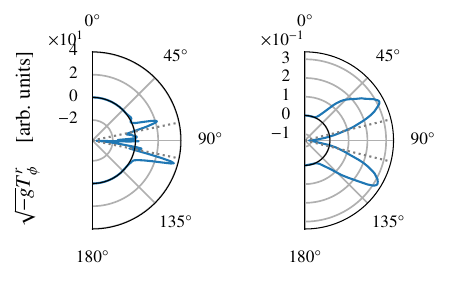}
\caption{Angular distribution of fluxes of rest mass $\sqrt{-g} \rho u^r$ (top panel), energy $\sqrt{-g} T^r_t$ (middle panel) and angular momentum $\sqrt{-g} T^r_\phi$ (bottom panel) through the spherical surfaces with radii $r/M = 9.38$ (outer edge of the initial torus) and $r/M = 21.55$ (a~few computational cells from outer boundary) from simulation for the Reissner-Nordström metric with $Q=1.09$. The cusp of the initial torus is located at $r_\text{cusp}/M=2.30$. The data are time averages of the simulation results from $t=10^4 t_g$ to $t=5 \times 10^4 t_g$.}
\end{figure}
\end{minipage}

\subsection{Angular momentum accretion rates}
\begin{figure}[!]
\centering
\includegraphics[width=0.49\columnwidth]{Figs/Q=1.02_RCUSP=2.65_10000_50000_jdot_insert}
\includegraphics[width=0.49\columnwidth]{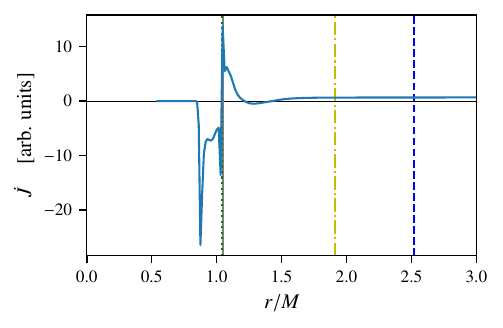}
\includegraphics[width=0.49\columnwidth]{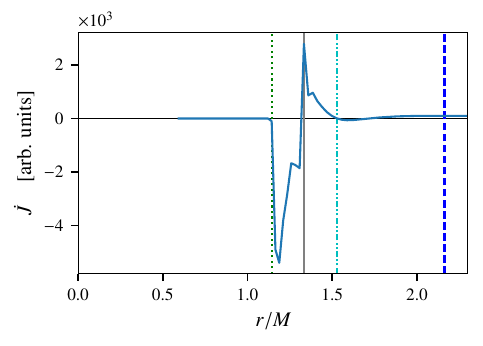}
\includegraphics[width=0.49\columnwidth]{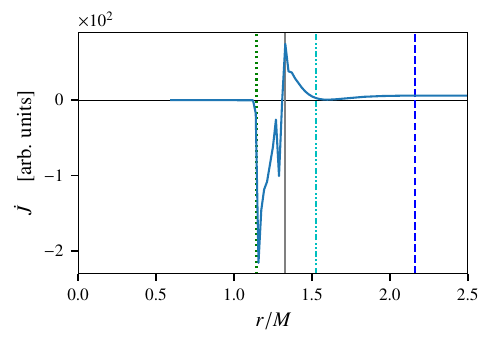}
\includegraphics[width=0.49\columnwidth]{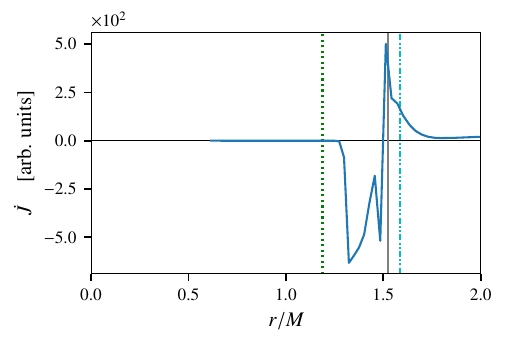}
\includegraphics[width=0.49\columnwidth]{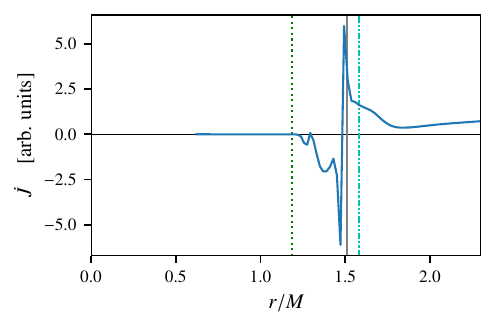}
\caption{The angular momentum accretion rate $\dot{J}$ as a function of the coordinate distance $r/M$ from the naked singularity in steady-state accretion in the Reissner-Nordström spacetime for $Q/M=1.02$ (top panels), $Q/M=1.07$ (middle panels) and $Q/M=1.09$ (bottom panels). The data are time averages of the results from $t=10^4 t_g$ to $t=5 \times 10^4 t_g$. Cusps of the initial tori are located at the right boundary of the plots. Green dotted lines correspond to zero-gravity sphere, yellow dash-dotted lines are the location of (outer) photon orbits. Dashed blue lines represent radii of marginally bound orbits. Cyan double dotted dashed lines are locations of maximum of Keplerian frequency $\Omega_\mathrm{K}$.}
\end{figure} 

\end{appendix}

\end{document}